\documentclass[11pt]{article}
\pdfoutput=1

\usepackage[usenames,dvipsnames,svgnames,table]{xcolor} 
\usepackage[obeyspaces,hyphens,spaces]{url}
\usepackage{jcapmod}

\usepackage[english]{babel}
\usepackage{blindtext}
\usepackage{amsmath, amsbsy, amssymb, amstext, amsthm,mathtools}
\usepackage{hyperref}
\usepackage{graphicx}
\usepackage{amsfonts}
\usepackage{fouridx}
\usepackage{cases}
\usepackage{bm}
\usepackage{bbm}
\usepackage{dsfont}
\usepackage{cleveref}
\usepackage{slashed}

\usepackage{setspace}

\usepackage{braket}

\usepackage{subcaption}

\definecolor{Blue}{rgb}{0.25, 0.41, 0.88}
\definecolor{Red}{rgb}{0.92,0.,0.}
\definecolor{darkorange}{rgb}{1.0,0.549,0.}
\definecolor{cobalt}{RGB}{44, 98, 120}
\definecolor{Mathematica1}{rgb}{0.368417, 0.506779, 0.709798}
\definecolor{Mathematica2}{rgb}{0.880722, 0.611041, 0.142051}
\definecolor{Mathematica3}{rgb}{0.560181, 0.691569, 0.194885}
\definecolor{Mathematica4}{rgb}{0.922526, 0.385626, 0.209179}
\definecolor{Mathematica5}{rgb}{0.528488, 0.470624, 0.701351}
\definecolor{Mathematica6}{rgb}{0.772079, 0.431554, 0.102387}
\definecolor{Mathematica7}{rgb}{0.363898, 0.618501, 0.782349}
\definecolor{Mathematica8}{rgb}{1, 0.75, 0}
\definecolor{Mathematica9}{rgb}{0.647624, 0.37816, 0.614037}
\definecolor{plotBlue}{RGB}{94, 130, 181}
\definecolor{plotRed}{RGB}{233, 85, 54}
\definecolor{plotGreen}{RGB}{142, 176, 50}
\definecolor{plotPurple}{RGB}{135, 120, 178}

\DeclareMathOperator*{\res}{res}
\DeclareMathOperator{\Li}{Li}
\DeclareMathOperator{\re}{Re}	
\DeclareMathOperator{\im}{Im}

\newcolumntype{C}[1]{>{\centering\let\newline\\\arraybackslash\hspace{0pt}}m{#1}}

\newcommand{\hodge}{\,\mathord{*}}

\makeatletter
\newlength{\apb@width}
\newcommand{\autoparbox}[2][c]{\settowidth{\apb@width}{#2}\parbox[#1]{\apb@width}{#2}}

\makeatother

\makeatletter
\newsavebox\myboxA
\newsavebox\myboxB
\newlength\mylenA

\newcommand*\xoverline[2][0.75]{
    \sbox{\myboxA}{$\m@th#2$}%
    \setbox\myboxB\null
    \ht\myboxB=\ht\myboxA%
    \dp\myboxB=\dp\myboxA%
    \wd\myboxB=#1\wd\myboxA
    \sbox\myboxB{$\m@th\overline{\copy\myboxB}$}
    \setlength\mylenA{\the\wd\myboxA}
    \addtolength\mylenA{-\the\wd\myboxB}%
    \ifdim\wd\myboxB<\wd\myboxA%
       \rlap{\hskip 0.5\mylenA\usebox\myboxB}{\usebox\myboxA}%
    \else
        \hskip -0.5\mylenA\rlap{\usebox\myboxA}{\hskip 0.5\mylenA\usebox\myboxB}%
    \fi}
\makeatother

\numberwithin{equation}{section}

\def\beq{\begin{equation}}
\def\eeq{\end{equation}}

\def\bea{\begin{eqnarray}}
\def\eea{\end{eqnarray}}

\def\beq{\begin{equation}}
\def\eeq{\end{equation}}
\def\bea{\begin{eqnarray}}
\def\eea{\end{eqnarray}}

\newcommand{\ud}{\mathrm{d}}
\newcommand{\lab}[1]{{\mathrm{#1}}}
\newcommand{\mb}[1]{{\mathbf{#1}}}
\newcommand{\minus}{{\scalebox{0.75}[1.0]{$-$}}}
\newcommand{\sminus}{{\scalebox{0.6}[0.85]{$-$}}}

\theoremstyle{definition}

\DeclareRobustCommand{\SkipTocEntry}[4]{}

\newcommand{\es}{\hspace{0.5pt}}
\newcommand{\ess}{\hspace{1pt}}

\definecolor{blue2}{cmyk}{1, 0.1, 0.1, 0.1}

\definecolor{pyBlue}{RGB}{31, 119, 180}
\definecolor{pyRed}{RGB}{214, 39, 40}
\definecolor{pyGreen}{RGB}{44, 160, 44}
\definecolor{pyBlue2}{RGB}{0, 111, 237}
\definecolor{pyRed2}{RGB}{224, 52, 36}

\begin{document}

\pagenumbering{roman}
\begin{titlepage}
\baselineskip=15.5pt \thispagestyle{empty}

\bigskip\

\vspace{1.5cm}
\begin{center}
{\fontsize{20}{22}\selectfont  {\bfseries Instanton Expansions and Phase Transitions}}
\end{center}
\vspace{0.42cm}
\begin{center}
{\fontsize{12}{18}\selectfont John Stout} 
\end{center}
\vspace{-0.2cm}
\begin{center}
\textit{Department of Physics, Harvard University, Cambridge, MA 02138, USA}
\end{center}

\vspace{1.2cm}
\hrule \vspace{0.25cm}
\noindent {\bf Abstract}\\[0.1cm]
	A central object in any axionic theory is its periodic potential, which is typically generated by instantons. The goal of this paper is to understand what physically happens to the theory when we lose control of the potential's instanton expansion. We argue, using the Yang-Lee theory of phase transitions, that the theory breaks down in the classic sense:  states become light.  However, these states are not necessarily light for all values of the axion and there can be large regions where the effective description remains valid. We find alternative expressions for the effective potential in terms of the properties of these light states, which remain useful even when the instanton expansion breaks down, and thus initiate a push beyond the lamppost of large instanton actions. 
	Most of these questions are motivated by the axionic Weak Gravity Conjecture, which we reformulate without reference to instanton actions. We also comment on its ability to constrain large-field axion inflation.
\vskip10pt
\hrule
\vskip10pt

\end{titlepage}

\thispagestyle{empty}
\setcounter{page}{2}
\begin{spacing}{1.03}
\tableofcontents
\end{spacing}

\clearpage
\pagenumbering{arabic}
\setcounter{page}{1}

\newpage

\section{Introduction}

	Quantum gravity famously abhors global symmetry, and the Weak Gravity Conjectures \cite{ArkaniHamed:2006dz,Brennan:2017rbf,Palti:2019pca} attempt to formalize just how pure that hatred is. They generally do so by asserting the existence of a set of states, coupled to each abelian gauge field in the theory, whose charge-to-mass ratio equals or exceeds that of an extremal black hole. If we try to restore the global symmetry by taking the gauge coupling to zero, these states must become light and the quantum gravitational theory fights back by destroying the validity of the effective description. These are thus conjectures about the behavior of quantum gravity near the edges of what is physically allowed. The goal of this paper is to understand what goes wrong when we try to restore a continuous shift symmetry.
	
	This the domain of the axionic Weak Gravity Conjecture (aWGC). Low-energy descriptions of string compactifications generically contain a plethora of periodic scalar fields \cite{Banks:2003sx, Svrcek:2006yi}. We will consider a single such field, $\varphi \sim \varphi + 2 \pi f$, whose field space circumference is measured by the axion decay constant~$f$. Typically, such a field classically enjoys a continuous shift symmetry $\varphi \to \varphi + \epsilon$ which is broken quantum mechanically by instantons. These non-perturbative effects generate an \emph{effective potential} of the schematic form
	\begin{equation}
		V_\lab{eff}(\varphi) = \sum_{\ell= 0}^{\infty} \Lambda_\ell^4 e^{-S_\ell}\left(1 - \cos \ell \varphi/f\right)\,. \label{eq:phiEffPot}
	\end{equation}
	This is best motivated in the dilute instanton gas approximation, where $S_\ell \approx |\ell| S_{1}$ roughly represents the action of instantons with topological charge $\ell$, and $\Lambda_\ell^4$ represents the contribution from fluctuations about them. Apparently, these symmetry-violating terms can be made arbitrarily small by taking the decay constant to be vastly super-Planckian, $f \gg M_\lab{pl}$, effectively restoring the global shift symmetry.

	On general grounds \cite{Banks:1988yz, Kallosh:1995hi,Banks:2010zn}, we do not expect this to be possible. This is borne out in examples~\cite{Banks:2003sx}, where it is very difficult to realize $f \gg M_\lab{pl}$ in a controlled manner. There are always states that become light or we lose control of the instanton expansion.  The axionic Weak Gravity Conjecture was originally proposed in \cite{ArkaniHamed:2006dz} as a way of formalizing the results of \cite{Banks:2003sx} and explaining the difficulty of attaining controlled string compactifications with super-Planckian decay constants. They argued, via the dimensional reduction of the Weak Gravity Conjecture for particles~\cite{delaFuente:2014aca,Heidenreich:2015nta} (see \cite{Brown:2015iha,Montero:2015ofa,Hebecker:2016dsw,Hebecker:2017wsu,Hebecker:2019vyf} for alternative attempts) that instanton actions are bounded by\footnote{There are different versions of this Weak Gravity Conjecture, varying in which instantons they constrain. In this work, we will only focus on bounds like (\ref{eq:awgc}) that predict vanishing instanton actions in the~$f/M_\lab{pl} \to \infty$ limit, and not work with a specific conjecture.}
	\begin{equation}
		S_\ell \lesssim {|\ell| M_\lab{pl}}/{f}\, \label{eq:awgc}
	\end{equation}
	in consistent theories of quantum gravity. This form mimics the analogous inequality for particles coupled to a $\lab{U}(1)$ gauge field, $m \lesssim |q| M_\lab{pl}\es e$, if we identify the mass $m$, gauge coupling $e$, and charge $q$ with the instanton's action $S_\ell$, inverse decay constant $f^{\sminus 1}$, and topological charge $\ell$, respectively.  As we take $f \gg M_\lab{pl}$, supposedly restoring the global symmetry, quantum gravity forces (\ref{eq:phiEffPot})'s higher harmonics to grow and cut down field range over which the potential is smooth. For this reason, there has been much excitement over whether or not the aWGC disallows large field inflation in quantum gravity; see \cite{Rudelius:2014wla,delaFuente:2014aca,Rudelius:2015xta,Brown:2015iha,Montero:2015ofa,Bachlechner:2015qja, Brown:2015lia,Junghans:2015hba,Heidenreich:2015wga, Palti:2015xra, Hebecker:2015rya,Hebecker:2015zss,Heidenreich:2015nta,Ibanez:2015fcv,Heidenreich:2019bjd} and the references therein.

	There are several problems with this story. First and foremost, it is exceedingly difficult to even define the action $S_\ell$ when the dilute instanton gas approximation is not applicable. How do we disentangle contributions to $S_\ell$ from those to the prefactor $\Lambda_\ell^4$? How do we separate the contributions due to single instantons from those of instantonic bound states \cite{Schafer:1996wv}? We could compute the action of some Euclidean saddle, but these have an extremely complicated relationship to the Fourier coefficients of (\ref{eq:phiEffPot}) when the actions are not large. These complications are all due to the simple fact that instantons are not physical objects like particles or branes, but are events which are introduced as a (sometimes very useful) tool for computation. 

	Neither is it clear from (\ref{eq:phiEffPot}) how quantum gravity supposedly responds when we attempt to restore the continuous shift symmetry. The other WGCs state things in very plain physical language: objects become light as we try to restore a global symmetry. The effective field theory breaks down because new degrees of freedom become relevant. The aWGC (\ref{eq:awgc}) instead predicts that the instanton expansion breaks down. But what exactly is wrong with needing to sum up a bunch of Fourier harmonics? There are plenty of well-behaved periodic functions with ill-behaved Fourier expansions. Would someone (or something) with sufficient computational might conclude that this limit is problematic? Or do we need to strengthen conditions like (\ref{eq:awgc}) to make it so? In general, we want to understand this strange asymmetry between the aWGC and its brethren.

	The goal of this paper is to thus answer the question: \emph{What goes wrong when an instanton expansion breaks down?} This is question is motivated by, but independent of, the aWGC. We will argue that the Fourier coefficients of the effective potential are unsuppressed when a state becomes light \emph{somewhere} along the axion's field space. The Fourier coefficients' asymptotic behavior is determined by the singularities of the potential, which are in turn related to closing energy gaps. We will argue for this via the Yang-Lee theory of phase transitions, which we will use to derive an alternative representation of the effective potential that is useful when the instanton expansion breaks down. This form makes it clear what data determines the effective potential in this limit, similar to how the properties of the Euclidean saddle approximately determine the potential when actions are large. Our arguments only rely on the analytic structure of the Euclidean partition function, and should thus be widely applicable. We will illustrate them in a variety of examples. 

	This perspective will allow us to unify our interpretation of the axionic Weak Gravity Conjecture with the others: they can all be thought of as a requirement that states must become light as we try to restore a global symmetry.
	Losing control of the instanton expansion is thus a signal that the effective field theory \emph{can} break down, somewhere along the axion's field space. However, as long as we stay away from these problematic points, the effective description remains useful. As we discuss in Section~\ref{sec:impl}, we can thus translate ill-defined criteria like (\ref{eq:awgc}) into well-defined bounds on the spectrum of the theory, with (\ref{eq:awgc2}) as an example.

	\paragraph{Outline} The plan of this paper is as follows: We introduce the general argument in Section~\ref{sec:ptFC}. We begin by reviewing the behavior of Fourier series in \S\ref{sec:asymp}, and then describe the Yang-Lee theory of phase transitions in \S\ref{sec:yangLee}. We combine these topics in \S\ref{sec:yangTopological}, where we specialize to periodic variables and find an expression for the effective potential that remains useful when the instanton expansion breaks down. We then present various examples in Section~\ref{sec:examples}, and comment on the implications of this perspective has for the axionic Weak Gravity Conjectures and large-field inflation in Section~\ref{sec:impl}. Finally, we present our conclusions in Section~\ref{sec:conclusions}. The appendices contain additional technical details that supplement the main text. 

\section{Phase Transitions and Fourier Coefficients} \label{sec:ptFC}

	Throughout this paper, we will focus on the theory of a single (pseudo-)scalar field $\theta$ with a compact field space $\theta \sim \theta + 2 \pi$, which we will call an axion. At lowest order in derivatives, this theory is described by the effective Lagrangian
 	\begin{equation}
		\mathcal{L} = -\frac{f^2}{2}\left(\partial \theta\right)^2 - V_\lab{eff}(\theta)\,, \label{eq:effLagrangian}
	\end{equation}
	where $f$ and $V_\lab{eff}(\theta)$ are the axion decay constant and effective potential, respectively. Our main goal is to identify the physics that controls this potential.

	Typically, the effective potential vanishes classically, but becomes nontrivial once we include quantum mechanical effects. These effects are usually non-perturbative in some parameter, though we will not assume that they have a particular origin. Instead, we will \emph{define} the effective potential as the vacuum energy of some UV theory at constant $\theta(t, \mb{x}) = \theta$ and take the axion to be non-dynamical. This may be extracted from the thermal partition function $\mathcal{Z}(\beta, \mathcal{V}, \theta)$ by taking the thermodynamic limit,
	\begin{equation}
		V_\lab{eff}(\theta) = -\lim_{\substack{\beta \to \infty \\ \mathcal{V} \to \infty}} \frac{1}{\beta \mathcal{V}} \log \mathcal{Z}(\beta, \mathcal{V}, \theta) \,.\label{eq:effectivePotentialDef}
	\end{equation}
	This partition function is computed at finite spatial volume $\mathcal{V}$ and (inverse) temperature $\beta$, and explicitly depends on the \emph{parameter}  $\theta$.\footnote{In this paper, we only consider $\theta$ as a classical parameter of a quantum mechanical theory, like a gauge theory's theta angle, and we will not allow it to quantum mechanically fluctuate. As discussed in \cite{Pimentel:2019otp}, the low-energy dynamics of the axion depends both on the potential defined by (\ref{eq:effectivePotentialDef}) as well as other corrections. Since, (\ref{eq:effectivePotentialDef}) is the main object of interest for the aWGC, we will focus only on it and leave the rest for future work.}  It can be defined as the Euclidean path integral over all fields in the theory on the spacetime cylinder, and is also a function of other parameters---like the gauge coupling or axion decay constant---whose dependence we suppress.

	Since the axion's field space is $2\pi$-periodic, the effective potential admits a Fourier expansion
	\begin{equation}
		V_\lab{eff}(\theta) = \sum_{\ell \in \mathbb{Z}} v_\ell\, e^{i \ell \theta}\,. \label{eq:fourierEffPot}
	\end{equation}
	Furthermore, we will assume that the potential (which is the ground state's energy density) is both real and integrable on $\theta \in [0, 2\pi)$, so that the Fourier coefficients obey $v_\ell = \bar{v}_{\sminus \ell}$ (the bar denotes complex conjugation) and fully determine the potential. 

	In this section, we first review some basic facts about the convergence of Fourier series in a single variable. We will show that the coefficients $v_\ell$ decay exponentially quickly as long as $V_\lab{eff}(\theta)$ is smooth, and identify what determines this rate of convergence. This decay becomes algebraic as soon as the potential develops a discontinuity in \emph{any} derivative \emph{anywhere} along the axion's field space, and the asymptotic behavior of these coefficients encodes both the nature and position of this discontinuity. We then review the Yang--Lee theory of phase transitions, which describes how these discontinuities are related to the appearance of light states and the breakdown of the effective theory. Finally, we use this to derive an alternative representation of the effective potential in terms of these light states that becomes useful when the topological (or instanton) expansion becomes unwieldy and unhelpful.

	\subsection{Asymptotic Behavior of Fourier Coefficients} \label{sec:asymp}

\begin{figure}
		\centering
		\includegraphics[]{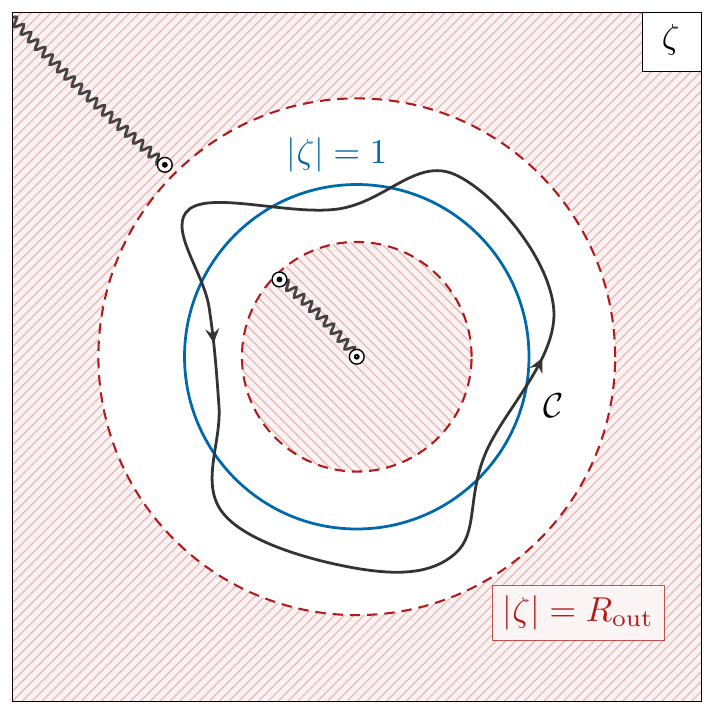}
		\caption{A function which is smooth and real along the unit circle $|\zeta| = 1$ is ``inversion symmetric'' and analytic within an annulus $\mathcal{A}$ (the white region) with radii $R_\lab{out} = 1/R_\lab{in}$. These radii are determined by the location of the nearest singularities, pictured here as black lines emanating from the origin and infinity. The function's Fourier/Laurent coefficients can be found by integrating along a contour $\mathcal{C}$ contained within the annulus. Their asymptotic behavior is thus determined by the size of this annulus.  \label{fig:laurent}}
	\end{figure}

	Let us first understand what determines how quickly the topological expansion (\ref{eq:fourierEffPot}) converges.\footnote{See \cite{Boyd:2001cfp} for an extremely readable introduction to the theory of polynomial and Fourier approximation, a subject very concerned with the convergence of such series. } It will be useful to complexify $\theta$ and extend the effective potential to the complex plane. To do so, we define the complex variable $\zeta \equiv e^{i \theta}$ and study the analytic continuation\footnote{This definition is only valid within the annulus $\mathcal{A}$, though it may be analytically continued beyond.}
	\begin{equation}
		V_\lab{eff}(\zeta) = \sum_{\ell \in \mathbb{Z}} v_\ell \zeta^\ell\,. \label{eq:complexExtension}
	\end{equation}
	Clearly, we recover the original potential if we restrict $V_\lab{eff}(\zeta)$ to the unit circle $|\zeta| = 1$, which we will call the \emph{physical domain}. If $V_\lab{eff}(\theta)$ is smooth for all $\theta$, this complex extension $V_\lab{eff}(\zeta)$ is analytic within an annulus
	\begin{equation}
		\mathcal{A} = \left\{\zeta \in \mathbb{C} \,\big|\, R_\lab{in} < |\zeta| < R_\lab{out}\right\},
	\end{equation} 
	whose size is determined by the nonanalyticities of the potential that are closest to the physical domain, see Figure~\ref{fig:laurent}.  Furthermore, since $V_\lab{eff}(\zeta)$ is a real-valued function on the unit circle, it is ``inversion symmetric'' $V_\lab{eff}(1/\bar{\zeta}) = \overline{V_\lab{eff}(\zeta)}$, and the radii  of $\mathcal{A}$ are reciprocals $R_\lab{out} = 1/R_\lab{in}$. 

	This complexification is quite useful, since the Fourier coefficients $v_\ell$ are the Laurent coefficients of $V_\lab{eff}(\zeta)$,
	\begin{equation}
		v_\ell = \frac{1}{2 \pi i} \int_\mathcal{C} \frac{\ud \zeta}{\zeta^{\ell +1}}\, V_\lab{eff}(\zeta)\,,
	\end{equation}
	where $\mathcal{C}$ is a counter-clockwise contour within $\mathcal{A}$. From elementary complex analysis, the Laurent series (\ref{eq:complexExtension}) only converges within $\mathcal{A}$, and thus the size of the annulus dictates how quickly the $v_\ell$ decay. For Taylor series, this is known as Darboux's theorem. Specifically, 
	\begin{equation}
		v_\ell \sim e^{-\mathcal{S} |\ell|} \,, \mathrlap{\qquad |\ell| \to \infty\,,}
	\end{equation}
	where we can define an ``asymptotic single instanton action'' $\mathcal{S} = \log R_\lab{out}$. However, note that this  nomenclature is only schematic---this definition does not rely on any sort of saddle-point approximation and $\ell\mathcal{S}$ is not, in general, the action of some stationary Euclidean configuration.
	
	We can now imagine changing $\mathcal{S}$ by varying the parameters of the theory that generates the effective potential---for instance, we could take the axion decay constant vastly super-Planckian, $f \gg M_\lab{pl}$, which we expect forces $\mathcal{S} \to 0$. As this happens, the width of the annulus $\mathcal{A}$ vanishes and there will be some nonanalyticity that impinges upon the physical domain, causing us to decide that we have ``lost control of the instanton expansion.''

	However, this does not necessarily mean that the potential becomes ill-behaved. Indeed, unless there is some instability that causes the effective potential to diverge, this ground state energy density $V_\lab{eff}(\theta)$ should still be integrable and continuous. Instead, this non-analyticity causes a derivative of the effective potential to develop a discontinuity \emph{somewhere} along the field space $\theta \in [0, 2\pi)$.  For instance, if the effective potential's $k$-th derivative becomes discontinuous at $\theta = a$, then the Fourier coefficients decay as\footnote{The extension to multiple discontinuities is trivial.}  
	\begin{equation}
		v_\ell = \frac{1}{2 \pi}\frac{e^{-i \ell a}}{(i \ell)^{k+1}} \,\,\underset{\theta=a}{\lab{disc}}\,\,\partial^k_\theta V_\lab{eff} + \mathcal{O}\big(\ell^{-k - 2}\big)\,. \label{eq:fourierAsymptotics}
	\end{equation}
	The asymptotic behavior of the Fourier coefficients are determined by the nature and location of its closest non-analyticity, and vice versa. Specifically, the nature of the discontinuity determines the Fourier coefficient's algebraic decay, while its location determines their oscillation frequency.

	The key point is that such singularities divide the physical domain $\theta \in [0, 2\pi)$ into different regions of analyticity, and that these regions define distinct phases of the theory~\cite{Goldenfeld:2018lop}. That is, \emph{the asymptotic behavior of the instanton expansion is determined by the nature and location of phase transitions that occur in the theory that generates the effective potential.} Such zero-temperature phase transitions generally occur when the energy gap between the vacuum and another state closes. We can thus phrase the ``breakdown'' of the instanton expansion as a statement about the Hilbert space of the theory, avoiding instantons altogether: \emph{as $\mathcal{S} \to 0$, an energy gap closes somewhere along the physical domain, signaling a breakdown of the effective theory.}

	It is important to note that because such a discontinuity can appear in a high-order derivative, this phase transition may occur at a point at which the potential is seemingly smooth and innocuous.\footnote{This is an explanation of the ``coherent instanton sum'' loophole (see, e.g. \cite{Heidenreich:2019bjd}) that allows one to simultaneously realize large-field inflation while satisfying the axionic Weak Gravity Conjecture.} This suggests that there may be an alternative way of describing this potential that does not require enormous amounts of data (i.e. millions of Fourier coefficients). Indeed, we will find it useful to instead describe the potential in terms of these singularities, whose properties have a physical interpretation. As we now review, the theory of phase transitions laid out by Yang and Lee in \cite{Yang:1952be,Lee:1952ig} provides this link.

	\subsection{The Yang--Lee Theory of Phase Transitions} \label{sec:yangLee}

		Historically, it was not clear whether the same partition function could describe two different phases of the same system---for instance, both the liquid and gaseous phases of water. This is related to the fact that the free energy for a system with finitely many degrees of freedom is analytic for real values of its parameters, and so there is no way for such a system to generate the singularities that would accompany such a transition. However, Lee and Yang \cite{Yang:1952be,Lee:1952ig} described how these singularities may emerge in the thermodynamic limit. We will quickly review their treatment of classical thermodynamic systems (see also, e.g. \cite{Reichl:2016msp,Bena:2005smo,Fisher:1967dta}) with a focus on details that will be relevant for the topological expansion discussed in the next section.

		Lee and Yang studied the $d$-dimensional classical lattice gas, a system of $N$ identical particles with mass $m$
		interacting pairwise via the ``hard core'' potential
		\begin{equation}
			V(|\mb{r}_{ij}|) = \begin{cases} 
									\infty & |\mb{r}_{ij}| < a \\
									\minus \epsilon_{ij} & a < |\mb{r}_{ij}| < b \\
									0 & b < |\mb{r}_{ij}|
								\end{cases}\,.
		\end{equation}
		This model is, in fact, mathematically equivalent to the classical Ising model. The grand partition function at coolness\footnote{Also known, more boringly, as the inverse temperature.} $\beta$, spatial volume $\mathcal{V}$, and chemical potential $\mu$, can be written as
		\begin{equation}
			\mathcal{Z}_{\beta, \mathcal{V}} (\zeta) = \sum_{n = 0}^{M} z_n \zeta^n\,, \label{eq:gasPartition}
		\end{equation}
		where we write it in terms of the fugacity
		\begin{equation}
			\zeta \equiv \left(\frac{m}{2 \pi \beta \hbar^2}\right)^{d/2} \!\!e^{\beta \mu}
		\end{equation}
		and the coefficients
		\begin{equation}
			z_n = \frac{1}{n!}\left[\,\prod_{i = 1}^{n} \int\!\ud^d \mb{r}_i\right] \exp\!\bigg(\!\minus \beta \sum_{i < j}^n V(|\mb{r}_{ij}|)\bigg)\,. \label{eq:configIntegral}
		\end{equation}
		Since these particles have an infinite hard core, only a finite number $M \sim \mathcal{V}/a^d$ may fit inside this box. The factor of $1/n!$ is necessary to avoid overcounting and plays an important role in determining (\ref{eq:gasPartition})'s region of convergence in the infinite volume ($M \to \infty$, $a$ constant) limit.

\begin{figure}
			\centering
			\begin{subfigure}{0.48\textwidth}
				\includegraphics[width=\textwidth]{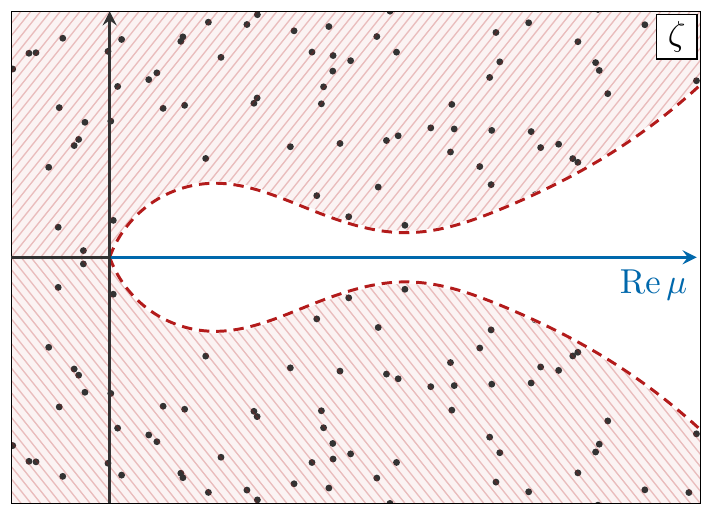}
				\caption{\label{fig:nophase}}
			\end{subfigure}
			\,\,\,\,
			\begin{subfigure}{0.48\textwidth}
				\includegraphics[width=\textwidth]{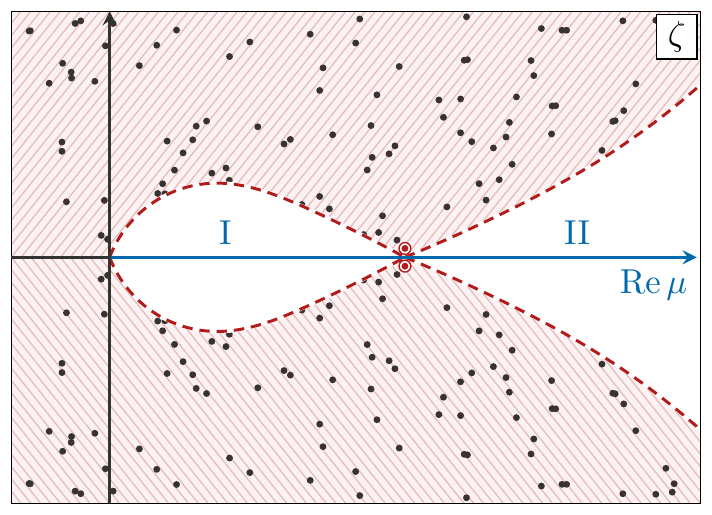}
				\caption{\label{fig:phase}}
			\end{subfigure}
			\caption{Partition functions zeros (some of which are shown here as black dots) form regions in the complex fugacity plane  (shown in white) in which we can analytically continue the pressure~$P_{\beta, \mathcal{V}}(\zeta)$. In the absence of a phase transition (left), this region of analyticity encompasses the entire physical domain and thus the pressure is smooth for all real values of the chemical potential. The system can also be tuned so that the zeros pinch the real axis (right), dividing the physical domain into two distinct phases (I and II), unrelated by analytic continuation. \label{fig:yangLeePT}}
		\end{figure}

		At finite volume, the partition function is an order $M$ polynomial in the fugacity $\zeta$. It is analytic everywhere in the complex plane, and is thus an entire function. We  denote its zeros as~$\{\zeta_k\}$, and thus the partition function can be factorized as
		\begin{equation}
			\mathcal{Z}_{\beta, \mathcal{V}}(\zeta) \propto  \prod_{k = 1}^M \left(1 - \frac{\zeta}{\zeta_k}\right)\,. \label{eq:factorizedPF}
		\end{equation}
		Because the coefficients $z_n$ are all real and positive, the zeros all lie away from the positive real axis and appear in complex-conjugated pairs, as illustrated in Figure~\ref{fig:nophase}. If we define the pressure at finite volume as
		\begin{equation}
			P_{\beta, \mathcal{V}}(\zeta) = \frac{1}{\beta \mathcal{V}} \log \mathcal{Z}_{\beta, \mathcal{V}}(\zeta) = \frac{1}{\beta \mathcal{V}} \sum_{k = 1}^{M} \log\left(1 - \frac{\zeta}{\zeta_k}\right) , \label{eq:pressureFinite}
		\end{equation}
		it is clear that this function must be smooth for real values of $\zeta$, as all singularities lie away from the real axis, and is thus analytic for physical values of the chemical potential.

		The situation can change in the thermodynamic limit. As $M \to \infty$, there may be zeros that impinge upon the real axis and divide the complex $\zeta$-plane into different regions of analyticity, as illustrated in Figure~\ref{fig:phase}. These regions of analyticity correspond to distinct phases of the system and the pressure (or one of its derivatives) will exhibit a singularity when we cross from one to another along the real axis. 

		In the thermodynamic limit, the zeros form a continuum that may be characterized by a density $\rho(\zeta)$. The pressure can then be written as
		\begin{equation}
			P_\beta(\zeta) = \lim_{\mathcal{V} \to \infty} \left[\frac{1}{\beta \mathcal{V}} \log \mathcal{Z}_{\beta, \mathcal{V}}(\zeta)\right] = \frac{1}{\beta} \int_{\mathbb{C}}\!\ud^2 \zeta' \, \rho(\zeta') \, \log \!\left(1 - \frac{\zeta}{\zeta'}\right)\,. \label{eq:thermoPressure}
		\end{equation}
		This form requires a little justification. It is clear that $\mathcal{Z}_{\beta, \mathcal{V}}(\zeta)$ is an entire function in $\zeta$ at finite $\mathcal{V}$. Its logarithm thus satisfies the Poincar\'{e}--Lelong equation \cite{Griffiths:1994pag}
		\begin{equation}
			\frac{1}{\pi i} \bar{\partial} \partial \log |\mathcal{Z}_{\beta}(\zeta)| = \{\mathcal{Z}_{\beta}\}\,, \label{eq:poincareLelong}
		\end{equation}
		where the right-hand side is the closed $(1,1)$-current associated with vanishing locus of $\mathcal{Z}_\beta(\zeta)$, and we use $\{\,\cdot\,\}$ to denote equivalence at the level of currents. Physically, (\ref{eq:poincareLelong}) has a simple interpretation. The zeros and poles of the meromorphic function behave as positive and negative point electric charges, respectively, and we may determine the function by solving (\ref{eq:poincareLelong}) (equivalently, Laplace's equation) supplemented with the appropriate boundary conditions.

		For instance, at finite $M$ the current takes the form
		\begin{equation}
			\{\mathcal{Z}_{\beta, \mathcal{V}}\} = \frac{1}{2 \pi i} \bar{\partial} \left[\,\sum_{k = 1}^{M} \frac{\ud \zeta}{\zeta - \zeta_k}\right],
		\end{equation}
		and we can solve (\ref{eq:poincareLelong}) to find
		\begin{equation}
			P_{\beta, \mathcal{V}}(\zeta) = \frac{1}{\beta \mathcal{V}} \log \mathcal{Z}_{\beta, \mathcal{V}} (\zeta) = g(\zeta) + \frac{1}{\beta} \sum_{k = 1}^{M} \frac{1}{\mathcal{V}} \log\left(1 - \frac{\zeta}{\zeta_k}\right). 
		\end{equation}
		Here $g(\zeta)$ is an entire function that is determined by the function's fall-off as $|\zeta| \to \infty$ (analogously, the boundary conditions of the Laplace problem). 

		From the Hadamard factorization theorem \cite{Boas:1954entire}, we know that the \emph{order} $\nu$ of $\mathcal{Z}_{\beta, \mathcal{V}}(\zeta)$ determines the degree of this function, $g(\zeta) \propto \zeta^\nu$ as $\zeta \to \infty$. The order can also be determined by the asymptotic behavior of the series coefficients
		\begin{equation}
			\nu = \limsup_{n \to \infty} \frac{n \log n}{\log(1/|z_n|)}\,.
		\end{equation}
		As long as the $z_n$ decay faster than factorially, $\nu = 0$ and $g(\zeta)$ is a constant factor which we can ignore. This depends on the behavior of the interaction potential: while the $1/n!$ from indistinguishability is helpful to ensure that the partition function is entire, this can be ruined by the ``configuration integrals'' in (\ref{eq:configIntegral}). However, one can show \cite{Ruelle:1999stat} that $\nu \leq 1$ if the interaction potential is \emph{stable}, and that $\nu = 0$ if it is \emph{superstable}.\footnote{The interested reader should consult \cite{Ruelle:1999stat} for definitions and an in-depth discussion of these stability conditions.} In what follows, we will assume that~$V(|\mb{r}_{ij}|)$ is superstable so that the order vanishes and we can safely ignore $g(\zeta)$.\footnote{This assumption also ensures that the density $\rho(\zeta)$ is such that the integral (\ref{eq:thermoPressure}) converges. Otherwise, a different Green's function would be needed, in analogy with the convergence-improving elementary factors that appear in the Hadamard factorization theorem. See \cite{Boas:1954entire} for more details.}

		As we approach the thermodynamic limit $\mathcal{V} \to \infty$, the number of zeros increases while the magnitudes of their individual contributions decrease. They thus form (usually linear) charge densities in the complex plane, and we may solve for the pressure (\ref{eq:thermoPressure}) using the two-dimensional Green's function.
		If these zeros pinch the real axis, then the pressure will develop a discontinuity in one of its derivatives at some physical value of the fugacity and the complex plane will be divided into separated regions of analyticity corresponding to distinct phases. Which derivative is discontinuous (i.e. the ``order'' of the phase transition) depends on how the zero density behaves near the real axis. For instance, if the $k$-th derivative suffers a discontinuity, the density vanishes as $\rho(\zeta) \propto (\im \zeta)^{k-1}$ as $\im \zeta \to 0$.

		In this section, we have seen how non-analyticities appear in the thermodynamic pressure of a classical system and how the pressure can be determined solely by knowing where the partition function vanishes (or perhaps becomes singular). Next, we will apply these ideas to the discontinuities that can appear in the quantum mechanically-generated potential for the axion $\theta$, and we will relate them to the presence of light states in the theory.	

	\subsection{Yang--Lee and the Topological Expansion} \label{sec:yangTopological}

		As we described at the beginning of this section, we define the effective potential (\ref{eq:effectivePotentialDef}) as the vacuum energy of the theory, which we may extract from the logarithm of the thermal partition function
		\begin{equation}
			\mathcal{Z}_{\beta}(\theta) = \lab{tr}\, e^{-\beta \mathcal{H}(\theta)}\, \label{eq:thermalPartitionFunction}
		\end{equation}
		in the low temperature limit $\beta \to \infty$. The goal of this section is to understand how the effective potential can be described solely in terms of the zeros and poles of this partition function, and to relate these analytic structures to the energy spectrum of the theory. 

		It will be helpful to summarize some basic facts about the analytic structure of $\mathcal{Z}_\beta(\theta)$. Since $\theta \to \theta + 2\pi$ is a gauge redundancy and the partition function is a single-valued\footnote{In principle, the partition function may pick up a phase as $\theta \to \theta + 2\pi$. As described in \cite{Cordova:2019jnf}, this only occurs if there is a level crossing and may be removed by the introduction of counterterms. We will assume that this has been done. Alternatively, we will only consider theories in the limit that a gap closes so that the partition function is fully $2\pi$-periodic.} in $\theta$, we can expand it into different topological sectors
		\begin{equation}
			\mathcal{Z}_\beta(\zeta) = \sum_{\ell \in \mathbb{Z}} z_{\ell} \zeta^\ell\,, \label{eq:pfLaurent}
		\end{equation}
		where we have extended the function to the complex plane by the map $e^{i \theta} \to \zeta$. As with the effective potential (cf. \S\ref{sec:asymp}), the partition function enjoys an inversion symmetry $\mathcal{Z}_\beta(1/\bar{\zeta}) = \overline{\mathcal{Z}_\beta (\zeta)}$ since it must be real on the unit circle $|\zeta| = 1$. It will be convenient to leverage this symmetry to factorize the partition function as
		\begin{equation}
			\mathcal{Z}_\beta(\zeta) = f(\zeta) \bar{f}(1/\zeta)\,, \label{eq:tpfF}
		\end{equation}
		where $f(\zeta) = \sum_{n=0}^{\infty} f_n \zeta^n$ is regular at the origin, though this factorization is not unique.

		As with the effective potential (cf. \S\ref{sec:asymp}), the Laurent expansion (\ref{eq:pfLaurent}) converges within an annulus about the unit circle $|\zeta| = 1$ whose inner and outer radii are reciprocals. We will first assume that this annulus covers the entire complex $\zeta$-plane, save for the points $\zeta = 0$ and $\zeta = \infty$. Then $f(\zeta)$ is an entire function and we only have to worry about the zeros of $\mathcal{Z}_\beta(\zeta)$, which we will also assume is order $0$. As we saw in the previous section, this is an assumption about the nature of the interactions in the theory that generates the partition function.

		We can motivate this assumption by considering a gauge theory partition function calculated in the dilute instanton gas approximation \cite{Coleman:1985rnk,Rajaraman:1982is}. The coefficient $z_\ell$ is the tunneling amplitude between the $0$-th and $\ell$-th topological sectors. When the instantons are well-localized, the amplitude can be approximated as
		\begin{equation}
			z_{\ell} \approx \sum_{n, \bar{n}} \frac{1}{n! \bar{n}!} \left(\beta \mathcal{K} e^{-S_{1}}\right)^{n + \bar{n}} \delta_{\ell, n - \bar{n}} \sim \frac{(\beta  e^{-S_1})^{|\ell|}}{|\ell|!}\,, \mathrlap{\qquad |\ell| \to \infty\,,}
		\end{equation}
		where $n$ ($\bar{n}$) denotes the number of $\pm1$-instantons contributing to the amplitude, $S_{1}$ is action of a $\pm1$-instanton,\footnote{Often one must also integrate over a scale quasi-zero mode and this changes the effective action of the contribution. That is, $S_1$ is not necessarily the action of some classical saddle. However, as long as these instantons remain relatively well-localized (for instance, if the gauge theory is higgsed), this argument should still work.} and $\mathcal{K}$ is contribution from fluctuations about the solution and other possible zero mode integrals. 
		Crucially, the indistinguishability of the instantons implies that there is an ``entropic suppression'' at large charge $\ell$. This is a representation of the fact that the $\ell$-instanton moduli space factorizes into $\ell$ copies of the single instanton moduli space when they are widely separated \cite{Dorey:2002ik}, weighted by a symmetry factor $1/\ell!$ to avoid overcounting. 

		In analogy with the classical lattice gas, this entropic suppression implies that the partition function is analytic almost everywhere in $\zeta$ as long as the  attractive interactions between instantons (mediated by fermions, for example) are not strong enough to ruin the large-$\ell$ asymptotics. For instance, this behavior does not qualitatively change in models where the dilute instanton gas approximation fails, as in the $\mathbb{C P}^{N}$ model or QCD~\cite{Schafer:1996wv,Diakonov:2002fq,Aguado:2002xf}. In both cases, fluctuations in the topological charge behave as $\log |z_\ell| \sim  \minus \ell^2$ \cite{Leutwyler:1992yt, KeithHynes:2008rw} and are roughly Gaussian, and so the function $f(\zeta)$ is not only entire but is also order $0$. In Appendix~\ref{app:asymp}, we provide an argument that this is the case in a wide class of theories, though it would be interesting to understand when and how this assumption can be violated.

		Working under the assumption that $f(\zeta)$ is entire, we can now consider the Poincar\'{e}-Lelong equation for the partition function at finite $\beta$. The current associated to its zero locus reads
		\begin{equation}
			\{\mathcal{Z}_{\beta}\} = \frac{1}{2 \pi i}\,\bar{\partial} \left[\, \sum_{k = 0}^{\infty} \frac{\ud \zeta}{\zeta - \zeta_k} + \sum_{k = 0}^{\infty} \left( \frac{\ud \zeta}{\zeta - \zeta_{\sminus k}} - \frac{\ud \zeta}{\zeta}\right)\, \right] \,.
		\end{equation}
		The additional $\ud \zeta/\zeta$ in the second sum should be interpreted as the image of the singularity at $\zeta = \infty$, and is necessary for the partition function to have the inversion symmetry that keeps it real along the unit circle. Similarly, this symmetry mandates that every zero which appears outside the unit circle, $\zeta_k$ for $k \geq 0$, must have a twin inside at $\zeta_{\sminus k} = 1/\bar{\zeta}_k$. We may then solve~(\ref{eq:poincareLelong}) to find 
		\begin{equation}
			\log \mathcal{Z}_\beta(\zeta) = g(\zeta) + \bar{g}(1/\zeta) + \sum_{k = 0}^{\infty} \log \left(1 - \frac{\zeta}{\zeta_k}\right) + \sum_{k = 0}^\infty \log \left(1 - \frac{\zeta_{\sminus k}}{\zeta}\right)\,, \label{eq:poincareSol}
		\end{equation}
		where $g(\zeta)$ is a polynomial whose degree is determined by the order $\nu$ of $f(\zeta)$. Assuming that $\nu = 0$, this is a constant which we can ignore.

		As before, these zeros condense into continuum distributions described by the densities $\rho_{\pm}(\zeta)$ in the low-temperature limit $\beta \to \infty$, so that the effective potential can be written as
		\begin{equation}
			V_\lab{eff}(\zeta) =  \int \!\ud \zeta' \, \rho_+(\zeta') \log \left(1 - \frac{\zeta}{\zeta'}\right) + \int \!\ud \zeta' \, \rho_{\sminus}(\zeta') \log \left(1 - \frac{\zeta'}{\zeta}\right)\,. \label{eq:effPotDensity}
		\end{equation}
		Here, $\rho_{+}$ ($\rho_-$) represents the zero density outside (inside) the unit circle. Interestingly, by expanding the logarithm and restricting to physical values $\zeta = e^{i \theta}$, 
		\begin{equation}
			V_\lab{eff}(\theta) = \sum_{\ell = 1}^{\infty} \left(\frac{1}{\ell} \int\frac{\ud \zeta}{\zeta^\ell} \,\rho_+(\zeta) \right) e^{i \ell \theta} + \sum_{\ell = 1}^{\infty} \left(\frac{1}{\ell} \int \!\ud \zeta \, \zeta^{\ell} \, \rho_{-}(\zeta)\right) e^{-i \ell \theta}\,,
		\end{equation}
		we see that the Fourier coefficients $v_\ell$ are nothing more than the multipole moments of these ``charge'' distributions. Intuitively, it is clear that as the zero density gets closer to the physical domain, more multipole moments are needed to describe it to a desired accuracy. Furthermore, we can focus entirely on the density outside of the unit circle, since the contribution from the inside can be recovered by complex conjugation.

		\begin{figure}
			\centering
			\includegraphics[]{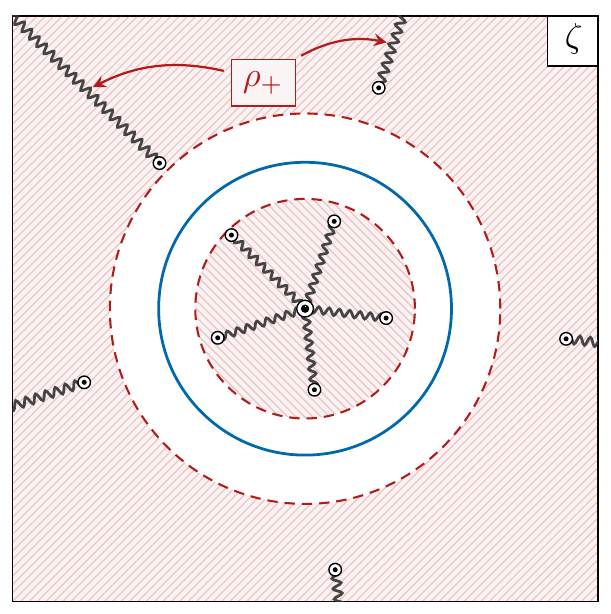}
			\caption{In this paper, we will focus on densities (pictured here as black lines) that extend radially outwards from the origin in the complex $\zeta$-plane. This will be the case in all of the examples we consider, and this restriction allows us to find simple expressions for the effective potential in terms of $\rho_+(\zeta)$, the zero density outside the physical domain. \label{fig:multi}}
		\end{figure}

		Where do these zeros come from? The partition function may be written as
		\begin{equation}
			\mathcal{Z}_\beta (\zeta) = e^{-\beta E_0(\zeta)} \bigg[1 + \sum_{i} e^{-\beta \Delta E_i(\zeta)}\bigg], \label{eq:pfZeroSum}
		\end{equation}
		where the $\Delta E_i(\zeta) \equiv E_i(\zeta) - E_0(\zeta)$ are the energy differences between the vacuum and the other states in the theory. 
		We see that the simplest set of zeros occurs when one of these energy gaps satisfies
		\begin{equation}
			\Delta E_i(\zeta_n) = -\frac{i \pi(2 n+1)}{\beta}\,, \mathrlap{\qquad n \in \mathbb{Z}\,.} \label{eq:zeroEq}
		\end{equation}
		In the thermodynamic limit $\beta \to \infty$, these zeros then form a continuum where the real part of the energy gap vanishes, ${\re \Delta E_{i}(\zeta) = 0}$. Clearly, this curve is parameterized by the imaginary energy gap $\epsilon_i(\zeta) = \im \Delta E_i(\zeta)$ and the density of zeros along it is
		\begin{equation}
			\rho_i(\zeta) \equiv \frac{1}{\beta}\frac{\ud n}{\ud \zeta} = -\frac{1}{2 \pi} \frac{\ud \epsilon_i}{\ud \zeta}\,. \label{eq:density}
		\end{equation}
		Eigenvalue repulsion implies that the curve ends at $\zeta_{i,*}$ where the both the real and imaginary parts of the gap vanish, $\Delta E_i(\zeta_{i, *}) = 0$. Since these energy differences generically behave asymptotically as  $\Delta E_i \propto -i \log \zeta$ as $|\zeta| \to \infty$ (cf. Appendix~\ref{app:asymp}), we expect $\epsilon_i$ to decrease towards $\minus \infty$ as we move away from $\zeta_{i,*}$.

		More complicated sets of zeros may arise when two or more states balance against the vacuum contribution in (\ref{eq:pfZeroSum}). This can only happen when the real energy gaps associated to each state vanish simultaneously, which generally occurs---if it occurs at all---only  at isolated points in the $\zeta$-plane.\footnote{There are highly-tuned models \cite{Biskup:2000gtl,Biskup2004:pfz} in which the curves defined by $\re \Delta E_i(\zeta) = 0$, $i = 1, 2, \dots$,  intersect tangentially. However, since this is not the generic situation, we will not consider it here.} Away from these so-called triple or higher coexistence points, the zeros behave as described above and we can treat each zero density independently. The point itself is measure zero, so its contribution to the potential does not survive the thermodynamic limit.

		Though there may be many curves that contribute to the effective potential, their importance is suppressed the further away they are from the unit circle. Since their contributions add linearly, we will restrict our focus to a single curve described by $\zeta(\epsilon)$, with $\epsilon \in (\minus\infty, 0]$ and $\zeta(0) = \zeta_*$. Since $\zeta_*$ sets the annulus of analyticity, we may identify $|\zeta_*| = e^{\mathcal{S}}$.  The effective potential is then
		\begin{equation}
			V_\lab{eff}(\zeta) = \frac{1}{2 \pi} \int_{\sminus \infty}^{0}\!\ud \epsilon \,\log \left(1 - \frac{\zeta}{\zeta(\epsilon)}\right) + \lab{c.c.} \label{eq:effPotLoc}
		\end{equation}
		Since $\zeta$ generally appears in the Hamiltonian in the combination $\log \zeta$ (i.e. it depends on $\theta$ and not $e^{i \theta}$) it will be convenient to define the coordinate
		\begin{equation}
			t(\epsilon) = \log \left(\frac{\zeta(\epsilon)}{\zeta_*}\right). \label{eq:tCurve}
		\end{equation}
		Integrating by parts, the effective potential may finally be written as
		\begin{equation}
			V_\lab{eff}(\theta) = -\frac{1}{2\pi}\int_{\mathcal{C}} \ud t \,\frac{\epsilon(t)}{1 - \zeta_* e^{t - i \theta}} + \lab{c.c.}
		\end{equation}
		where the curve $\mathcal{C}$ goes from $t = 0$ to $\infty$ through the path defined by (\ref{eq:tCurve}). 

		This form is particularly useful because, in the cases we consider, $\zeta(\epsilon)$ describes a curve that moves radially outwards from $\zeta_*$ (as pictured in Figure~\ref{fig:multi}) and thus
		\begin{equation}
			V_\lab{eff}(\theta) = -\frac{1}{2 \pi}\int_{0}^{\infty}\!\ud t\, \frac{\epsilon(t)}{1 - \zeta_* e^{t - i \theta}}\, + \lab{c.c.} \label{eq:effPotAlt}
		\end{equation}
		The integrand is exponentially suppressed for large values of $t$, so only the behavior of $\im \Delta E(t) = \epsilon(t)$ near $t = 0$ (where the gap fully vanishes) is important. If we replace $\epsilon(t)$ with its expansion,
		\begin{equation}
			\epsilon(t) \sim 2\pi \sum_\alpha \frac{\epsilon_\alpha\, t^\alpha}{\Gamma(\alpha+1)}\,, \mathrlap{\quad\qquad t \to 0\,,} \label{eq:epsExp}
		\end{equation}
		with the exponents $\alpha$ positive but not necessarily integer, we see that the effective potential admits a natural decomposition
		\begin{equation}
			V_\lab{eff}(\theta) = \sum_\alpha \epsilon_\alpha \Li_{\alpha+1}(e^{i\theta}/\zeta_*) + \lab{c.c.} \label{eq:polyLogExp}
		\end{equation}
		in terms of polylogarithms. These polylogarithms provide a natural basis for the effective potential when the Fourier expansion becomes unwieldy, in the same sense that (co)sinusoids are a natural basis for the potential when the dilute instanton gas approximation is applicable.

		As we discussed in the previous section, the effective potential becomes discontinuous when a density of zeros impinges upon the physical domain, and such a discontinuity is associated with a closing energy gap. This can be easily seen from (\ref{eq:polyLogExp}), since $\Li_{k}(\zeta)$ has a branch cut emanating from $\zeta = 1$. Specifically, if $\zeta_* = e^{i \delta}$, then the $\ell$-th derivative $\partial_\theta^\ell \Li_{\ell+1}(e^{i \theta}/\zeta_*)$ is discontinuous at $\theta = \delta$, and so the leading behavior of $\Delta E(t)$ near $\zeta_*$ determines the nature of this discontinuity.

		When compared to the topological expansion (\ref{eq:fourierEffPot}), the representation (\ref{eq:effPotAlt}) provides more efficient way of describing the effective potential, especially away from the dilute instanton gas limit when higher harmonics are not exponentially suppressed. One only needs to specify the location $\zeta_*$ at which the states become light and their number---which affects the leading exponent $\alpha$ in (\ref{eq:epsExp})---in order to get a approximate description of the effective potential.  Of course, that is not to say that~(\ref{eq:effPotAlt}) is a cure-all---one must still identify and calculate the properties of the states that are becoming light as one moves around the axion's field space. It does, however, provide a physical description of what is happening in the theory beyond ``the instanton expansion becomes poorly-controlled,'' and allows one to describe the effective potential using relatively little data that may be calculated in a controlled way.

		We conclude by noting that we can not rule out the possibility that the thermal partition function of a general theory has a more exotic analytic structure than we assumed here. For instance, the Kaluza-Klein reduction of a complex scalar field, discussed in \S\ref{sec:extra}, provides an example where the partition function is not entire in $\zeta$. As we describe there, gapless modes contribute poles to the partition function that restrict the annulus of analyticity. The methods outlined here still apply, as we are able to determine that $g(\zeta)$ is a constant directly from the path integral. The main utility of these assumptions is to relate singularities in the effective potential to the spectrum of the theory. In more exotic cases we may need to relax our conclusion that ``the instanton expansion breaks down when states become light'' to the more tautological ``the instanton expansion breaks down when there is a phase transition.'' It would be interesting to understand if there are non-pathological theories where this is the case.
	
\section{Some Illustrative Examples} \label{sec:examples}

	In the previous section, we provided a general argument relating the asymptotic behavior of the effective potential's Fourier coefficients to the behavior of light states in the theory. The potential's Fourier expansion cannot converge exponentially quickly if there is a vacuum degeneracy for some value of $\theta$. We will now confirm this behavior in a number of examples and explicitly identify the responsible light states. 

	In \S\ref{sec:toy}, we study what is perhaps the simplest model with interesting nonperturbative dynamics: the particle on a circle in an electromagnetic field. The model shares many features with higher-dimensional quantum field theories and has the benefit of being completely solvable. In~\S\ref{sec:extra}, we study a model in which the effective potential is not generated by instantons, but is instead the Casimir energy induced by a tower of Kaluza-Klein modes. Finally, in \S\ref{sec:other}, we explain how this story applies to potentials that arise in Yang-Mills theory and  string compactifications.
	
	\subsection{A Low-Dimensional Model} \label{sec:toy}

		We will begin with a toy quantum mechanical model: the particle on a circle in an electromagnetic field. This theory is described by the action
		\begin{equation}
			S_\lab{1d} = \int\!\ud t\, \left(\frac{1}{2 g^2} \dot{A}^2  - \theta \dot{A} - V(A)\right)\,, \label{eq:pocToy}
		\end{equation}
		along with the identification $A \sim A+1$, where the potential $V(A)$ represents an electric field $E = -\ud V$ placed along the circle and $\theta$ encodes the magnetic flux passing through it. 

		We should think of this as a ``minisuperspace'' reduction\footnote{In fact, this model (with vanishing potential) is a sector of two-dimensional $\lab{U}(N)$ gauge theories on a spacetime cylinder \cite{Lawrence:2012ua}. One may also derive a similar effective description of $\lab{SU}(N)$ gauge theories on $\mathbb{R}\times \lab{S}^3$ by restricting to the gauge field zero modes, see e.g. \cite{Luscher:1982ma,Luscher:1983gm,vanBaal:1988qm,vanBaal:1992xj,vandenHeuvel:1994ah}.} of the prototypical quantum field theory used to study nonperturbative dynamics,
		\begin{equation}
			S_\lab{4d} = \int \left(-\frac{1}{4 g^2}\, \lab{tr}\, F\wedge  \!\hodge F + \frac{\theta}{16 \pi^2} \, \lab{tr}\, F \wedge F\right) + S_\lab{cm}\,, \label{eq:4dGauge}
		\end{equation}
		where $F = \ud A + A \wedge A$, $A$ is a non-Abelian gauge field with a topologically non-trivial field space, and $S_\lab{cm}$ is the action of matter charged under $A$. Intuitively, the kinetic term $\dot{A}^2$ mimics $\lab{tr}\, F\wedge \hodge F$, $\theta \dot{A}$ mimics $ \theta \, \lab{tr}\, F \wedge F$, and the potential $V(A)$ mimics the contribution from charged matter~$S_\lab{cm}$. We will discuss this relation in more detail in \S\ref{sec:yangMills}.

		It will be convenient to restrict to a simple cosine potential $V(A) = \Lambda(1 - \cos 2 \pi A)$, as the resulting Schr\"{o}dinger equation reduces to the well-studied Mathieu equation. Our toy model's Hamiltonian is then
		\begin{equation}
			\mathcal{H} = \frac{g^2}{2}\left(p_A + \theta\right)^2 + \Lambda\left(1 - \cos 2 \pi A\right)\,, \label{eq:toyHam}
		\end{equation}
		where the canonical momentum is $p_A = \dot{A}/g^2 - \theta$. The identification $A \sim A +1$ forces the wavefunction to be \emph{invariant} under discrete translations
		\begin{equation}
			\psi(A) = \psi(A+1)\,. \label{eq:toyShifts}
		\end{equation}
		Conceptually, it will be  convenient to work instead with a ``gauge transformed'' wavefunction
		\begin{equation}
			\tilde{\psi}(A) = e^{-i \theta A} \psi(A), \label{eq:1dbc}
		\end{equation}
		so that the Hamiltonian becomes that of a one-dimensional, perfectly sinusoidal crystal,
		\begin{equation}
			\mathcal{H}_\lab{c} = \frac{g^2}{2} p_A^2 + \Lambda \left(1 - \cos 2 \pi A\right)\,. \label{eq:crystalHam}
		\end{equation}
		For a crystal of infinite extent, Bloch's theorem guarantees that the energy eigenstates are quasi-periodic in $A$,
		\begin{equation}
			\mathcal{H}_\lab{c} \psi_{n, \kappa}(A) = E_{n}(\kappa) \psi_{n, \kappa}(A) \qquad \text{where} \qquad \psi_{n, \kappa}(A+1) = e^{i \kappa} \psi_{n, \kappa}(A).
		\end{equation}
		In the condensed matter literature, $\kappa$ is called the quasimomentum and is said to take values in the (first) Brillouin zone, here $\kappa \in [-\pi, \pi)$. Furthermore, the energy eigenfunctions $\psi_{n, \kappa}(A)$ are called the Bloch waves of the $n$'th energy band, which are the Mathieu functions for this perfectly sinusoidal crystal. For an infinite crystal, there is a continuum of energy eigenfunctions parameterized by $\kappa$ for each integer $n$. However, the boundary condition (\ref{eq:1dbc}) removes this continuum and forces us to consider only $\kappa = \theta$. These energy eigenvalues are then related to the Mathieu characteristics\footnote{The Mathieu characteristic can be defined as the $e_n(q, \theta)$ for which the recurrence relation ${(e_n(q, \theta) - 4(m - \theta/2 \pi)^2)c_{m} + q (c_{m+1} + c_{m-1}) = 0}$ has a minimal (normalizable) solution. These minimal solutions are labeled by the integer $n$. This is a computationally convenient definition, as the Mathieu characteristics are then well-approximated by the eigenvalues of the tridiagonal matrix encoded by this relation.}
		\begin{equation}
			E_{n}(\lambda, \theta) = \frac{\pi^2 g^2}{2} \left(\frac{4 \lambda}{\pi} + e_{n}(2 \lambda/\pi, \theta)\right)\,.
		\end{equation}
		where we have defined the dimensionless coupling $\lambda \equiv \Lambda/(2 \pi g^2)$. The effective potential is then the minimum among these, $V_\lab{eff}(\theta) = E_0(\lambda, \theta)$.

\begin{figure}
			\centering
			\makebox[\textwidth][c]{\includegraphics[width=1\textwidth]{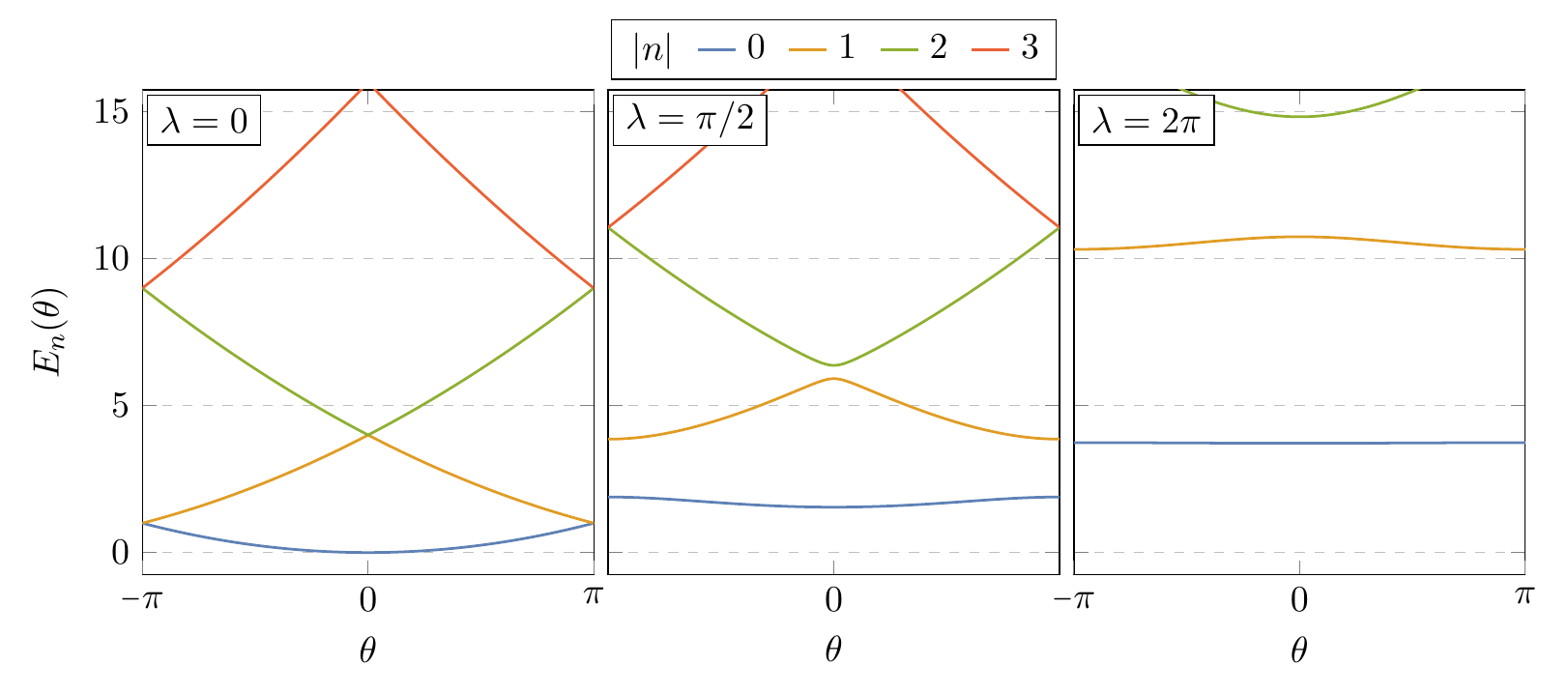}}
			\caption{The lowest energy eigenvalues (in units of $\pi^2 g^2/{2}$) of the toy model (\ref{eq:toyHam}) for various values of the potential height $\lambda$. The effective potential is the lowest level, shown in blue. \label{fig:toyEnergies}}
		\end{figure}

		When $\lambda = 0$, this potential is most easily represented by the minimum among a set of harmonic branches,
		\begin{equation}
			V_\lab{eff}(\theta) =\min_{n \in \mathbb{Z}}  \frac{g^2}{2} \left(\theta - 2 \pi n\right)^2\,,
		\end{equation}
		while for $\lambda \gg 1$ this can be computed via an instanton calculation \cite{Asorey:1983hd}. In this case, the thermal partition function is dominated by instantons of the form
		\begin{equation}
			\bar{A}(t) = \pm \frac{2}{\pi} \arctan e^{2 \pi g \sqrt{\Lambda}(t - t_0)}
		\end{equation}
		with action
		\begin{equation}
			S_1 = 4 \sqrt{\frac{2 \lambda}{\pi}}\,. \label{eq:instAction}
		\end{equation}
		The effective potential is then well-approximated by an exponentially suppressed cosine,
		\begin{equation}
			V_\lab{eff}(\theta) = \frac{\pi^2 g^2}{2} \left(\sqrt{\frac{8 \lambda}{\pi}} - \frac{16 \sqrt{2 }}{\pi^2} \lambda^{1/2} e^{-4\sqrt{2 \lambda/\pi}} \cos \theta + \dots \right). \label{eq:toyInst} 
		\end{equation}
		In the latter case, each $\pm1$-instanton is well-localized with width $\mathcal{O}(\lambda^{\sminus 1/2})$, while in the former they are completely delocalized and the Fourier coefficients $v_\ell = g^2 (-1)^\ell/\ell^2$ decay algebraically. From (\ref{eq:fourierAsymptotics}) and the discussion of \S\ref{sec:yangTopological}, we can predict that an energy gap closes at $\theta = \pm \pi$ as $\lambda \to 0$. 
		We confirm this in Figure~\ref{fig:toyEnergies}, where we plot the first few energy levels as functions of $\theta$ for illustrative values of $\lambda$.

		We can understand the behavior of the effective potential by studying the analytic structure of the thermal partition function 
		\begin{equation}
			\mathcal{Z}_\beta(\lambda, \zeta) = \sum_{n = 0}^{\infty} \exp\left(-\beta E_{n}(\lambda, \zeta)\right)
		\end{equation}
		in the low-temperature $\beta \to \infty$ limit. It will be easiest to first focus on the case where $\lambda = 0$, where the partition function is just the Jacobi theta function,
		\begin{equation}
			\mathcal{Z}_\beta(0, \zeta) =  \frac{1}{\sqrt{2 \pi g^2 \beta}} \sum_{\ell \in \mathbb{Z}} q^{\ell^2} \zeta^\ell 
		\end{equation}
		where $q = e^{-1/(2 g^2 \beta)}$, and may be written using the Jacobi triple product as 
		\begin{equation}
			\mathcal{Z}_{\beta}(0, \zeta) = \frac{1}{\sqrt{2 \pi g^2 \beta}}\prod_{n = 0}^{\infty} \left(1 - q^{2n+2}\right)\left(1 + q^{2n+1}\zeta\right)\left(1 + q^{2n+1}\zeta^{\sminus 1}\right)\,. \label{eq:jacobiTriple}
		\end{equation}
		From this product form, we can infer that the partition function is entire with zeros at
		\begin{equation}
			\zeta_k = -\exp\left(\frac{2 k +1}{2 g^2 \beta}\right)\,,\mathrlap{\qquad k \in \mathbb{Z}\,.}
		\end{equation}
		The zeros all lie along the negative real axis and those with $k \geq 0$ lie outside the unit circle. Furthermore, their zero-temperature density (\ref{eq:density}) impinges upon the physical domain $|\zeta| = 1$ and decays linearly with $\zeta$, $\rho(\zeta) = \minus g^2/\zeta$.

		\begin{figure}
				\centering
				\includegraphics[width=\textwidth]{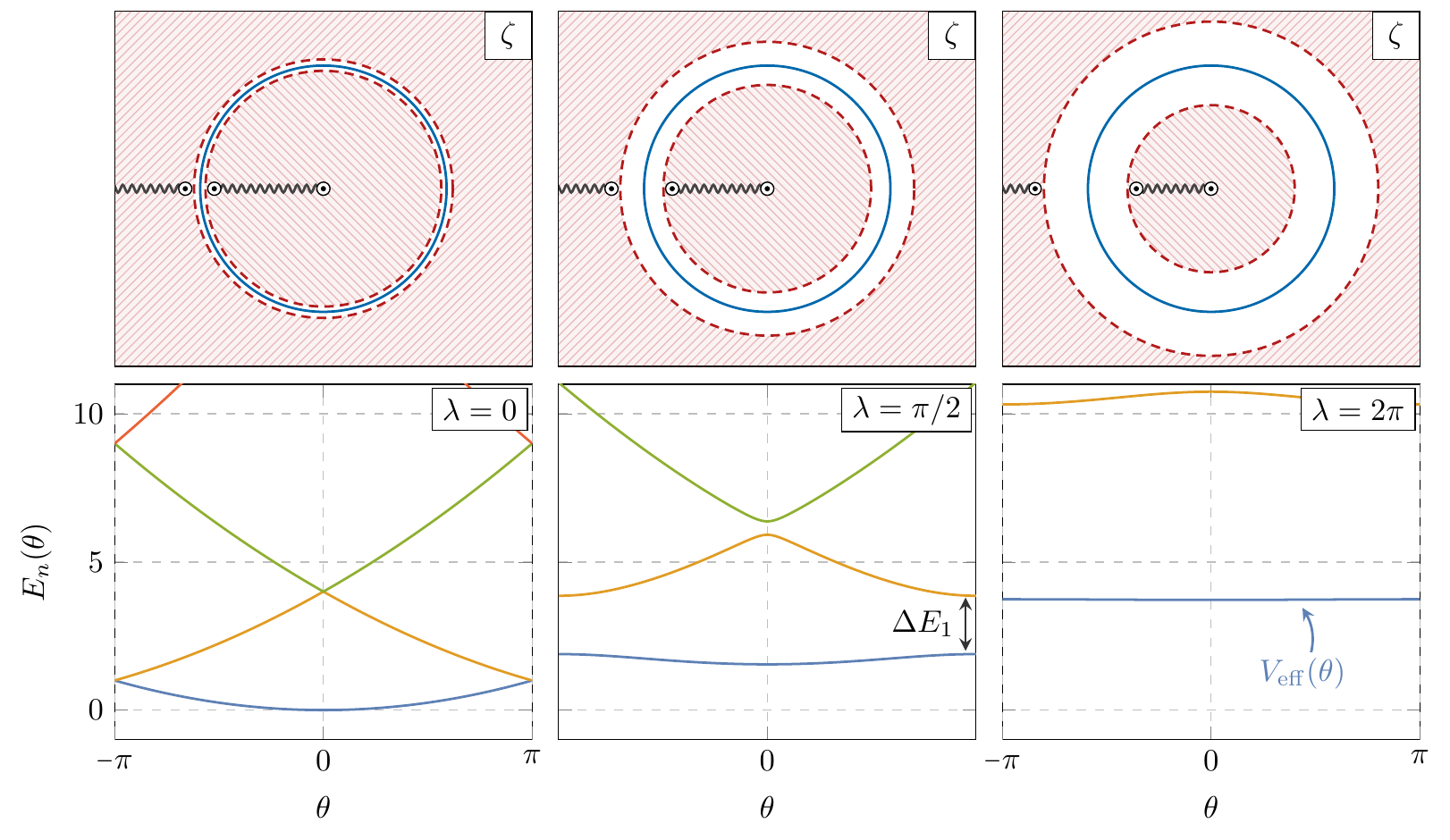}
				\caption{A schematic representation of the partition function zeros (top) as a function of $\lambda$, plotted against the energy eigenvalues as functions of $\theta$ (bottom). As the density of zeros pinches the physical domain $|\zeta| =1$ (left), the effective potential (in blue) develops a cusp at $\theta = \pm \pi$. As the zeros move away (center and right), this cusp smooths out and the gap  $\Delta E_1$ grows. \label{fig:zeroPlots}}
		\end{figure}

		We can also infer this zero density by studying the energy differences
		\begin{equation}
			\Delta E_{k}(\zeta) = 2 \pi g^2 k \left(\pi k + \lab{arg}\, \zeta\right) - 2 \pi g^2 i k \log |\zeta|
		\end{equation}
		as a function of $\zeta$. The only energy gap that fully closes (even in the complex plane) is $\Delta E_1(\zeta)$, and its real part vanishes for real, negative $\zeta$. In terms of the imaginary part $\epsilon_1 \equiv \im \Delta E_1$, this traces out a curve
		\begin{equation}
			\zeta(\epsilon_1) = -\exp\left(\!-\frac{\epsilon_1}{2 \pi g^2}\right)\,,
		\end{equation}	
		which intersects the physical domain at $\zeta_* = -1$.

		Introducing the coordinate $t = \log(\zeta(\epsilon_1)/\zeta_*)$ (so that $\epsilon_1 = -2 \pi g^2 t$) and using (\ref{eq:effPotDensity}) and (\ref{eq:effPotAlt}), the effective potential is
		\begin{align}
			V_\lab{eff}(\theta) =  - g^2 \int_{0}^{\infty} \! \frac{\ud t\,t}{1 + e^{t - i \theta}} + \lab{c.c.} =  g^2 \left[\lab{Li}_{2} \big(\minus e^{i \theta}\big) + \lab{Li}_{2}\big(\minus e^{-i \theta}\big)\right]
		\end{align}
		which is simply a complicated way of writing
		\begin{equation}
			V_\lab{eff}(\theta) =  \min_{\ell \in \mathbb{Z}} \frac{g^2}{2} \left(\theta-2 \pi \ell\right)^2  -\frac{g^2 \pi^2}{6} \,. \label{eq:periodicQuadratic}
		\end{equation}
		As expected, we recover the effective potential up to an overall shift in the vacuum energy.

\begin{figure}
			\centering
			\includegraphics[]{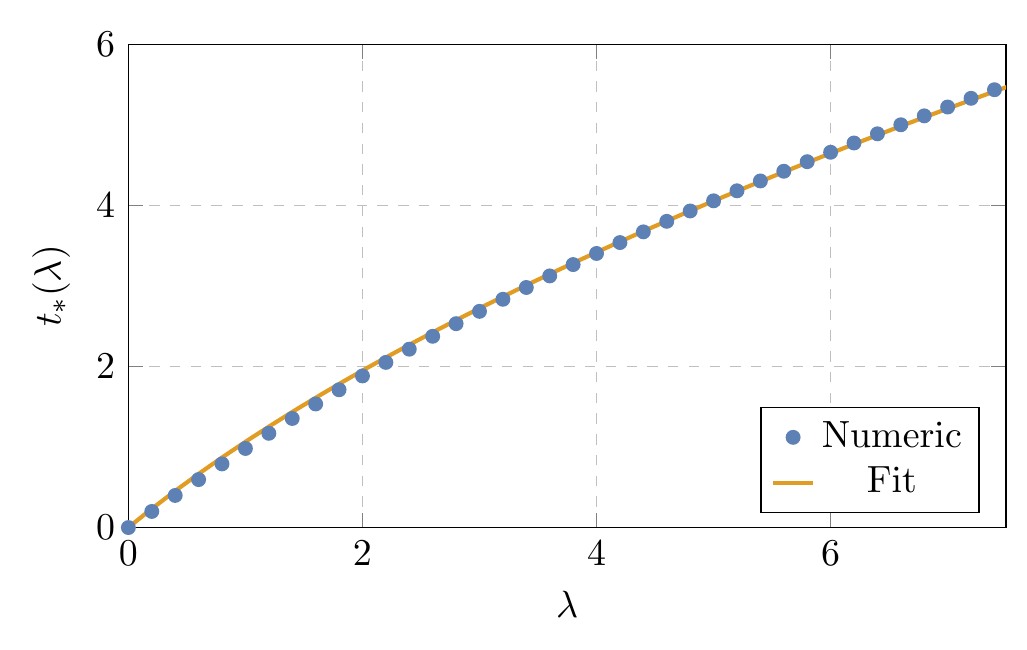}
			\caption{We plot the numerically determined $t_*(\lambda)$ (blue marks) against a simple fit $4 \sqrt{2/\pi} (\sqrt{\lambda + \sqrt{\pi}} - \pi^{1/4})$ (solid yellow). This matches the expected large $\lambda$ asymptotics from the instanton expansion, (\ref{eq:toyInst}). This simple fit overestimates the slope at small $\lambda$ by about $20\%$, for which $t_* \sim \lambda$, as expected from degenerate perturbation theory. \label{fig:tsFit}}
		\end{figure}

		As we increase $\lambda$ and make it harder for $A$ to tunnel around its field space, this density of zeros moves away from the unit circle, as shown in Figure~\ref{fig:zeroPlots}. For small $\lambda$, eigenvalue repulsion tells us that
		\begin{equation}
			\Delta E_1(\zeta) \sim 2 \pi g^2 \sqrt{\lambda^2 - \log^2(-\zeta )}\,. \label{eq:energyGapToy}
		\end{equation}
		In fact, the form 
		\begin{equation}
			\epsilon_1(t) = 2 \pi g^2 \sqrt{t(t + 2 t_*(\lambda))}
		\end{equation}
		proves a good approximation for $\epsilon_1$ at moderate $\lambda$. Here, we have introduced $t_* = \log |\zeta_*(\lambda)| \approx \mathcal{S}$, a function of $\lambda$ that goes as $t_*(\lambda) \sim \lambda$ when $\lambda \gg 1$ and $t_*(\lambda) \sim 4\sqrt{2\lambda/\pi}$ as $\lambda \gg 1$, which measures how far away from the unit circle the nearest singularity is.  We can confirm its behavior by numerically determining where $\Delta E_1(t_*) = 0$ and plotting it against a simple fit $t_*(\lambda) \approx 4\sqrt{2/\pi}(\sqrt{\lambda + \sqrt{\pi}} - \pi^{1/4})$, shown in Figure~\ref{fig:tsFit}. Importantly, the asymptotic behavior of the Fourier coefficients are dominated by the physics of a single instanton when $\lambda \gg 1$, while for $\lambda \ll 1$ they are dominated by the physics of eigenvalue repulsion. Both extremes offer a simple, some expansion point around which we can understand the behavior of the effective potential.

		From this density, we can exactly calculate the Fourier coefficients
		\begin{equation}
			v_\ell = g^2 (-1)^{\ell+1} e^{-\ell t_*}\int_{0}^{\infty}\!\ud t \, \sqrt{t(t+ 2t_*)} \,e^{- \ell t} =  \frac{g^2 (-1)^{\ell+1}}{\ell^2} (\ell t_*) K_{1}(\ell t_*)\,.
		\end{equation}
		For $\ell t_* \ll 1$, we have
		\begin{equation}
			v_\ell = \frac{g^2 (-1)^{\ell+1}}{\ell^2} \left( 1 + \frac{1}{2} \left[\log\left(\frac{ \ell t_*}{2}\right) + \gamma_{\textsc{E}} - \frac{1}{2}\right] \ell^2 t_*^2 + \dots\right), \label{eq:exp1}
		\end{equation}
		where $\gamma_{\textsc{E}}$ is the Euler-Mascheroni constant, while for $\ell t_* \gg 1$,
		\begin{equation}
			v_\ell = \frac{g^2 (-1)^{\ell+1}}{\ell^{3/2}}\sqrt{\frac{ \pi t_*}{2}} e^{-\ell t_*} \left(1 + \frac{3}{8 \ell t_*} + \dots\right)\,. \label{eq:exp2}
		\end{equation}
		As we explain in Appendix~\ref{app:genDen}, it is best to perform a small-$t_*$ expansion for the entire function $V_\lab{eff}(\theta)$ instead of the individual Fourier coefficients, in which case we find that it admits a series expansion in terms of polylogarithms and their derivatives,
		\begin{align}
				 V_\lab{eff}(t_*, \theta) &\sim g^2\Li_{2}({e^{i \theta}}/{\zeta_*}) + g^2 t_* \Li_{1}(e^{i \theta}/\zeta_*) \nonumber \\
				&+ \frac{g^2t_*^2}{4} \left(\left(1 + 2 \gamma_{\textsc{e}} + 2 \log \tfrac{t_*}{2}\right) \Li_{0}(e^{i \theta}/{\zeta_*}) - 2\Li_{0}^{\scriptscriptstyle{(1)}}(e^{i \theta}/\zeta_*)\right) + \mathcal{O}(t_*^3) + \lab{c.c.}
		\end{align}
		where $t_* = \lambda + \mathcal{O}(\lambda^2)$, $\zeta_* = e^{t_*}$, and $\Li_{n}^{\scriptscriptstyle{(1)}}(\zeta) = \partial_n \Li_{n}(\zeta)$. We can thus avoid awkwardly patching together the two expansions (\ref{eq:exp1}) and (\ref{eq:exp2}) for the Fourier coefficients by expanding the effective potential directly in the energy gap $t_*$.

		\subsubsection*{Fractional Instantons and the Topological Expansion} \label{sec:toyFrac}

		Before we move onto the next section, it will be useful to deform our toy model (\ref{eq:toyHam}) and consider
		\begin{equation}
			\mathcal{H}_\lab{c} = \frac{g^2}{2} p_A^2 + \Lambda(1 - \cos 2 \pi N A)\,,. \label{eq:nToyHam}
		\end{equation}
		This theory enjoys a $\mathbb{Z}_N$ discrete shift symmetry $A \to A + 1/N$. As described in \cite{Unsal:2012zj,Gaiotto:2017yup}, this model is qualitatively similar to $\lab{SU}(N)$ Yang-Mills, where this shift symmetry mimics the gauge theory's $\mathbb{Z}_N$ center symmetry. Like in the double well, we expect that the Hilbert space is divided into nearly-degenerate subspaces whose energies are split by nonperturbative effects. This implies that there are $N$ nearly degenerate ``vacuum'' states that form a ``vacuum family''~\cite{Gabadadze:2002ff}.  In fact, since the $\mathbb{Z}_N$ shift symmetry enjoys a mixed 't Hooft anomaly with time-reversal symmetry at $\theta = \pi$ \cite{Gaiotto:2017yup}, these ``vacuum'' states must interchange as we shift $\theta \to \theta + 2\pi$ and we thus expect that the effective potential has a cusp, even if the height of the potential $\Lambda$ is large and we expect a sort of instanton expansion to be useful. It will be useful to see how this works in more detail.

		The potential in (\ref{eq:nToyHam}) has $N$ degenerate minima and so there are fractional instantons that tunnel between them,
		\begin{equation}
			\bar{A}(t) = \frac{j}{N} \pm \frac{2}{\pi N} \arctan e^{2\pi g N \sqrt{\Lambda}(t - t_0)}
		\end{equation}
		with $j = 0, 1, \dots, N -1\,$, action $S_1/N$ (\ref{eq:instAction}), and topological charge $\pm 1/N$. The thermal partition function can then be approximated as a sum over these fractional instantons,
		\begin{equation}
			\mathcal{Z}_{\beta}(\theta) \propto \sum_{n, \bar{n} = 0}^{\infty}\!\frac{(\beta \mathcal{K})^{n+\bar{n}}}{n! \bar{n}!} e^{-(n+\bar{n})S_1/N + i (n - \bar{n})\theta/N} \left(\,\sum_{\ell \in \mathbb{Z}} \delta_{n - \bar{n}, \ell} \right)\,. \label{eq:fracInstSum}
		\end{equation}
		Here, we use $\mathcal{K}$ to denote the contribution from Gaussian fluctuations about each instanton, and the term in parenthesis ensures that we restrict to configurations that come back to themselves, as required by the Euclidean path integral's boundary conditions. Using $\sum_{\ell \in \mathbb{Z}} \delta_{n - \bar{n}, \ell} \propto \sum_{k = 0}^{N-1} e^{-2\pi i(n - \bar{n}) k/N}$, we can rewrite this as 
		\begin{equation}
			\mathcal{Z}_\beta(\theta) \propto \sum_{k = 0}^{N-1} \sum_{n, \bar{n} = 0}^{\infty}\!\frac{(\beta \mathcal{K})^{n+\bar{n}}}{n! \bar{n}!} e^{-(n+\bar{n})S_1/N + i (n - \bar{n})(\theta+ 2\pi k)/N} = \sum_{k = 0}^{N-1} e^{-\beta E_k(\theta)} \label{eq:fractionalThermal}
		\end{equation}
		with the energies
		\begin{equation}
			E_k(\theta) = \frac{\pi^2 g^2}{2} \left[ N \sqrt{\frac{8 \lambda}{\pi}} - \frac{16 \sqrt{2}}{\pi^2	} \lambda^{1/2} e^{-S_1/N} \cos\!\left(\frac{\theta - 2\pi k}{N}\right) + \dots \right]\,.
		\end{equation}
		Note that we can interpret the effect of the constraint in (\ref{eq:fracInstSum}) as introducing $N$ non-trivial instanton phases which we must sum over to impose the correct boundary conditions. The effective potential is then the minimum among these branches $V_\lab{eff}(\theta) = \min_k E_k(\theta)$, as in~Figure~\ref{fig:qcdEnergies}. 

		If we analytically continue the energies into complex $\zeta = e^{i \theta} = e^{t + i \delta}$, with $\delta \in (-\pi, \pi]$, the energy gaps $\Delta E_{k}(\zeta) = E_k(\zeta) - E_0(\zeta)$ may be written as
		\begin{equation}
			\Delta E_{k}(\zeta) = 2 \alpha  \sin\!\big(\tfrac{\pi k}{N}\big) \Big[ \sin\!\big(\tfrac{\pi k - \theta}{N}\big) \cosh\!\big(\tfrac{t}{N}\big) + i \cos\!\big(\tfrac{\pi k - \theta}{N}\big) \sinh \!\big(\tfrac{t}{N}\big) \Big]\,,
		\end{equation}
		with $\alpha = 8 \sqrt{2} g^2 \lambda^{1/2} e^{-S_1/N}$. In the low-temperature limit, the thermal partition function is zero whenever the real part of one of these gaps vanishes, which only happens when $\theta = \pi$ and $k = 1$. We are then able to recover the effective potential from the imaginary part of this energy gap,
		\begin{equation}
			\epsilon(t) = 2 \alpha \sin\!\big(\tfrac{\pi}{N}\big)\,\sinh \!\big(\tfrac{t}{N}\big)\,.
		\end{equation}
		While we can exactly evaluate (\ref{eq:effPotAlt}) with this density in terms of hypergeometric functions, it is more convenient to confirm that this recovers $V_\lab{eff}(\theta) = -\alpha \cos \!\big(\tfrac{\theta}{N}\big)$ with $\theta \in (-\pi, \pi]$ by explicitly comparing their Fourier coefficients,\footnote{Note that our derivation of (\ref{eq:effPotAlt}) relied on our assumption that the thermal partition function behaved as an order $0$ entire function as $|\zeta|\to \infty$. Even though the approximation (\ref{eq:fractionalThermal}) grows as $\log \mathcal{Z}_\beta(\zeta) \sim \zeta^{1/N}$ as $|\zeta| \to \infty$, we find that (\ref{eq:effPotAlt}) still works.} 
		\begin{align}
			v_\ell &=    \frac{\alpha (-1)^\ell }{\pi \csc \!\big(\tfrac{\pi}{N}\big) }\,\int_{0}^{\infty}\!\ud t \, \sinh\!\big(\tfrac{t}{N}\big) e^{-|\ell| t} = -\frac{\alpha}{2 \pi} \int_{-\pi}^{\pi}\!\ud \theta \, \cos \!\big(\tfrac{\theta}{N}\big) \,e^{-i \ell \theta} = \frac{N \alpha}{2\pi \csc \!\big(\tfrac{\pi}{N}\big) } \frac{(-1)^\ell}{N^2 \ell^2 - 1}\,,
		\end{align}
		which perfectly agree for $\ell \neq 0$.

		As expected, the degeneracy at $\theta = \pi$ forces the Fourier coefficients to asymptotically decay algebraically, $v_\ell \propto \ell^{-2}$. Interestingly, the partition function can be computed using a well-controlled fractional instanton expansion, even though a light state forces the topological expansion to converge slower than exponentially. The fractional instanton expansion remains useful because it explicitly retains the light states that dictate the asymptotic behavior of the Fourier coefficients $v_\ell$. In this simple model, it was clear that such fractional instantons existed and that they provided a useful way to organize our perturbative expansion. However, if we did not know about them and instead brutishly tried to compute the coefficients $v_\ell$ directly, we might be convinced that we would need to compute the multitudinous corrections from multi-instanton configurations.
	
		\vspace{1pt}
\subsection{An Extra-Dimensional Model} \label{sec:extra}

		We will now consider an somewhat orthogonal example: the dimensional reduction of a $(d+2)$-dimensional complex scalar field $\Phi(x^\mu, y)$ along a circle of circumference $R$, where $x^\mu  \in \mathbb{R}^{1,d}$ and $y \in [0, R)$. This theory is often called \emph{extranatural inflation} \cite{ArkaniHamed:2003wu,Heidenreich:2015nta} and is described by the action
		\begin{equation}
			S = -\int\!\ud^{d+2} x\,\left[ \partial_\mu \Phi \,\partial^\mu \bar{\Phi} + \partial_y \Phi \partial_y \bar{\Phi} + m^2 |\Phi|^2 \right],
		\end{equation}
		where we introduce the theta angle by requiring that the scalar field is quasi-periodic along the compact direction, $\Phi(x^\mu, y+ R) = e^{i \theta} \Phi(x^\mu, y)$. We will find that the vacuum energy of this theory depends on $\theta$ and that, as in the previous example, its discontinuities are related to the presence of light states in the spectrum.

		It is convenient to expand $\Phi(x^\mu, y)$ in terms of its Kaluza-Klein modes,
		\begin{equation}
			\Phi(x^\mu, y) = \frac{1}{\sqrt{R}} \sum_{\ell \in \mathbb{Z}} \Phi_\ell(x^\mu) \exp\left[i \big(\theta+2 \pi \ell\big)y/R\right]\,,
		\end{equation}
		so that the action can be written as a sum over a tower of $(d+1)$-dimensional complex scalar fields,
		\begin{equation}
			S = \sum_{\ell \in \mathbb{Z}} \int\!\ud^{d+1}x\, \left[- \partial_\mu \Phi_{\ell}\, \partial^\mu \bar{\Phi}_{\ell} - \left[m^2 +   m_{\textsc{kk}}^2(\theta + 2 \pi \ell)^2\right]  \Phi_{\ell} \bar{\Phi}_{\ell}\right], \label{eq:extraAction}
		\end{equation}
		where we have introduced the Kaluza-Klein mass scale $m_{\textsc{kk}}^2 = R^{-2}$.
		
		We can compute the thermal partition function via the Euclidean path integral by decomposing each field $\Phi_\ell(x^\mu)$ into Fourier and Matsubara modes, which we denote by $\mb{k}$ and $\omega_{n} = 2 \pi n/\beta$, respectively. The partition function then factorizes into a product of Gaussian integrals which can be readily evaluated to yield\footnote{We used a primed index $n'$ to indicate that we exclude the Matsubara zero mode $n = 0$ from the product.} 
		\begin{equation}	
			\mathcal{Z}_\beta(\theta) = \prod_{\ell, \mb{k}} \left[T \prod_{n_\ell} (\omega_{n_\ell}^2 + E^2_{\ell,\mb{k}}(\theta))^{-1} \prod_{n'} \omega_n^2\right]. \label{eq:matsubaraPF}
		\end{equation}
		Here, we have introduced the dispersion relation
		\begin{equation}
			E_{\ell, \mb{k}}^2(\theta) = \mb{k}^2 + m^2 + m_{\textsc{kk}}^2 \left(\theta + 2 \pi \ell\right)^2\,.
		\end{equation}
		By summing over Matsubara modes \cite{Laine:2016hma}, we may rewrite this as
		\begin{equation}
			-\frac{1}{\beta \mathcal{V}} \log Z_{\beta}(\theta) = \sum_{\ell \in \mathbb{Z}} \int\!\!\frac{\ud^d k}{(2 \pi)^d} \left[E_{\ell, \mb{k}}+ \beta^{\sminus 1}\log \left(1 - e^{-\beta E_{\ell, \mb{k}}(\theta)}\right)\right] \label{eq:thermalSummed}
		\end{equation}
		where $\mathcal{V}$ is the spatial volume. 

		The low-temperature limit of (\ref{eq:thermalSummed}) yields the effective potential, which reduces to a sum over the vacuum energies associated to each Fourier mode,
		\begin{equation}
			V_\lab{eff}(\theta) = \sum_{\ell \in \mathbb{Z}} \int\!\!\frac{\ud^d k}{(2 \pi)^d} \, E_{\ell, \mb{k}}(\theta) = \int\!\!\frac{\ud^d k}{(2 \pi)^d} \sum_{\ell \in \mathbb{Z}} \sqrt{\mb{k}^2 + m^2 + m_{\textsc{kk}}^2(\theta + 2 \pi \ell)^2}\,.
		\end{equation}
		This can be evaluated using standard techniques \cite{Ambjorn:1981xw}
		\begin{equation}
			V_\lab{eff}(\theta) =  -\frac{1}{4 \pi}\frac{m^{d+2}}{m_{\textsc{kk}}}\left[ \frac{\Gamma\big(\minus \tfrac{d+2}{2}\big)}{(4 \pi)^{{d}/{2}}} + \frac{8}{(2 \pi)^{{d}/{2}}} \sum_{n = 1}^{\infty} \frac{K_{d/2+1} (m R n) \cos n \theta}{(m R n)^{{d}/{2}+1}} \right] \label{eq:effPotExtra}
		\end{equation} 
		We will ignore the $\theta$-independent constant throughout.

		In the limit that the higher-dimensional mass vanishes $m \to 0$, the effective potential again reduces to a polylogarithm,
		\begin{equation}
			V_\lab{eff}(\theta) = -\frac{m_{\textsc{kk}}^{d+1}}{\pi^{d/2 +1}} \Gamma\big(\tfrac{d+2}{2}\big) \Li_{d+2}(e^{i \theta}) + \lab{c.c.} \label{eq:lowmassLimit}
		\end{equation}
		which exhibits a discontinuity in its $(d+1)$-th derivative at $\theta = 0$. Our general discussion in \S\ref{sec:yangTopological} suggests that an energy gap closes there and we see from (\ref{eq:extraAction}) that, indeed, the Kaluza-Klein zero mode $\Phi_0(x^\mu)$ becomes massless in this limit.

		\begin{figure}
			\centering
			\includegraphics{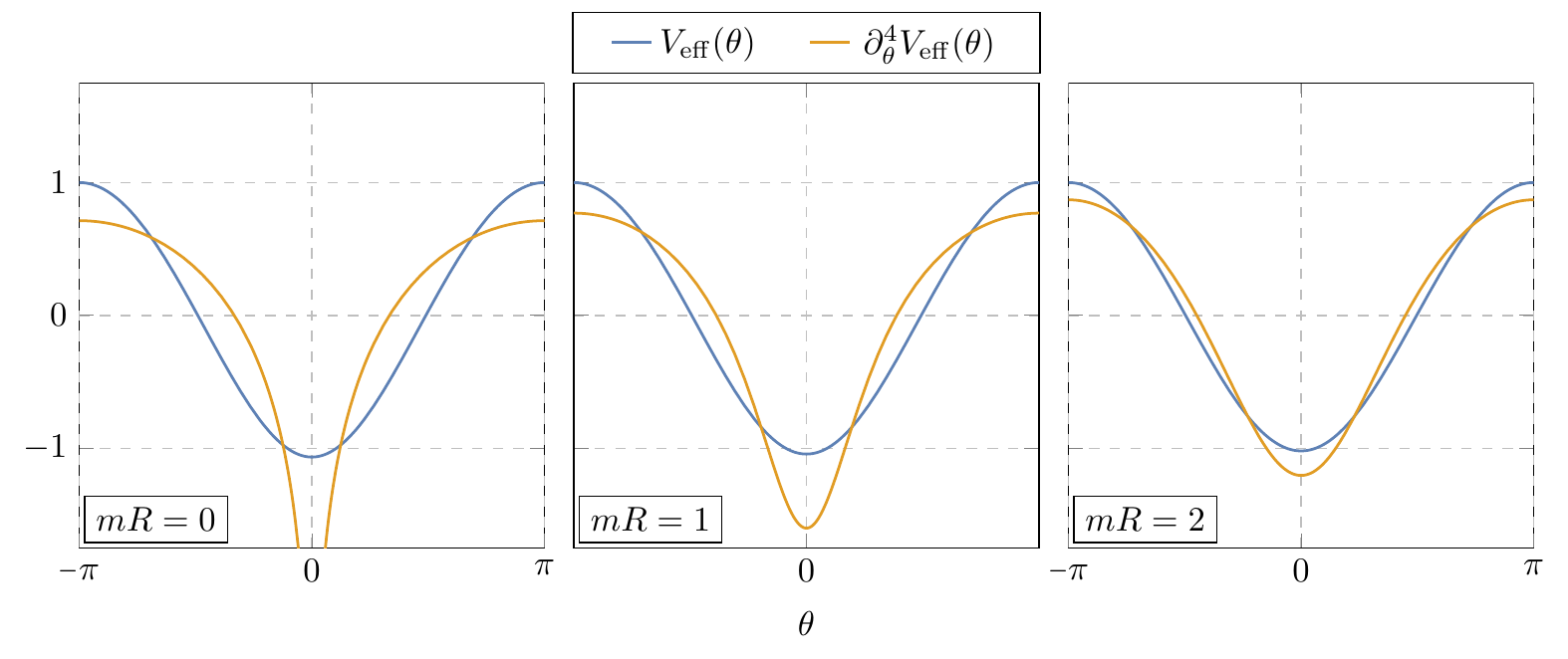}
			\caption{The effective potential (\ref{eq:extraNatural3d}) and its fourth derivative for various values $m R$, normalized so that $V_\lab{eff}(\pi) = 1$. While the effective potential barely changes as we dial $m R$ to zero, its fourth derivative at $\theta = 0$ diverges, forcing the topological expansion to converge algebraically. \label{fig:extra}}
		\end{figure}

		The effective potential (\ref{eq:effPotExtra}) simplifies into a finite sum of polylogarithms for odd $d$. For example, if $d = 3$, then
		\begin{align}
			V_\lab{eff}(\theta) &= -\frac{3 m_{\textsc{kk}}^4}{4 \pi^2}\left[\Li_5\!\big(e^{i \theta}/\zeta_*\big) + (m R) \Li_4\!\big(e^{i \theta}/\zeta_*\big) + \frac{(m R)^2}{3} \Li_3 \!\big(e^{i \theta}/\zeta_*\big) + \lab{c.c.}\right] \nonumber \\
			&= -\frac{3 m_{\textsc{kk}}^4}{2 \pi^2} \sum_{\ell = 1}^{\infty} \left[\frac{1}{\ell^5} + \frac{m R}{\ell^4} + \frac{(m R)^2}{3 \ell^3}\right] e^{-\ell m R} \cos \ell \theta \label{eq:extraNatural3d}
		\end{align}
		which has a discontinuous third derivative at $\theta = 0$ as $m R \to 0$, somewhat like a third-order phase transition. As we show in Figure~\ref{fig:extra}, the potential hardly changes (aside from an overall scaling) as a function of $m R$, even though the coefficients of the topological expansion (\ref{eq:extraNatural3d}) no longer exponentially decay. We reiterate that the convergence of the topological expansion depends on how smooth the effective potential is for all $\theta$, and that the loss of this exponential decay merely signals that \emph{some} derivative of the potential is blowing up. This, however, does not necessarily imply that the potential is badly-behaved in this limit.

	\subsubsection*{From the Poles}

		We will now relate the effective potential (\ref{eq:effPotExtra}) directly to the analytic structure of the thermal partition function, as described generally in \S\ref{sec:yangTopological}. We will be able to see explicitly how the discontinuity at $\theta = 0$ is related to a vanishing energy gap.

		It will be convenient to first rewrite the modified Bessel function in (\ref{eq:effPotExtra}) as
		\begin{equation}
			z^{-\nu} K_{\nu}(z) = \frac{\sqrt{\pi} e^{-z}}{2^\nu \Gamma\big( \nu + \tfrac{1}{2}\big)}\int_{0}^{\infty}\!\ud x\, e^{-z x} \left[x(x+2)\right]^{\nu - \frac{1}{2}}\,.
		\end{equation}
		We can then perform the geometric sum over $n$ to write the effective potential in the general form (\ref{eq:effPotDensity}) with $\zeta_* = e^{m R}$ and density
		\begin{equation}
			\epsilon(t) =  -\frac{2 \pi m_{\textsc{kk}}^{d+1}}{\Gamma\big(\tfrac{d+3}{2}\big)} \left[\frac{t(t+2 m R)}{4 \pi}\right]^{\frac{d+1}{2}}\,. \label{eq:extraDensity}
		\end{equation}
		This density is negative, and from the general discussion of \S\ref{sec:yangTopological} we expect that the effective potential is formed by the \emph{poles} of the partition function instead of its zeros.

		Indeed, if we evaluate the thermal partition function in terms of the Matsubara modes (\ref{eq:matsubaraPF}), we see that these poles occur whenever
		\begin{equation}
			\omega_{n_\ell}^2 + E_{\ell, \mb{k}}^2(\zeta) = 0.
		\end{equation}
		This condition can only be satisfied for $\theta \in [-\pi, \pi)$ by the $\ell = 0$ sector. We find that there are two solutions for each Matsubara mode, one inside and one outside the unit circle, and that they all lie on the positive real axis. Explicitly, the poles outside the unit circle can be labeled by their Matsubara number $n \in \mathbb{Z}$ and their $d$-dimensional momentum $\mb{k}$, 
		\begin{equation}
			\zeta(n, \mb{k}) = \exp\left(R \sqrt{\mb{k}^2 + (2 \pi n/\beta)^2 + m^2}\right),
		\end{equation}
		and so the effective potential can be written as
		\begin{equation}
			V_\lab{eff}(\theta) = \lim_{\beta, \mathcal{V} \to \infty} \sum_{\mb{k},\, n} \frac{1}{\mathcal{V} \beta} \log\left(1 - e^{i \theta}/\zeta(n, \mb{k})\right) + \lab{c.c.}
		\end{equation}
		As discussed previously, these partition function poles contribute to (\ref{eq:poincareLelong}) with an opposite sign as compared to its zeros.

		In the zero-temperature limit, the poles again form a continuum on the real axis. We may thus convert the above sum into an integral over a pole density,
		\begin{equation}
			V_\lab{eff}(\theta) = \int\!\frac{\ud^{d+1}k}{(2 \pi)^{d+1}} \, \log\left(1 - e^{i \theta - R \sqrt{k^2 + m^2}}\right)
		\end{equation}
		where we have identified $k_0 = \omega_n$. We can perform the angular integrals and define $t = R \sqrt{\mb{k}^2 + m^2} - m R$ so that, upon integrating by parts, we can rewrite the effective potential in the form (\ref{eq:effPotAlt}) with density (\ref{eq:extraDensity}).

	\subsubsection*{From the Casimir Energy}

		Finally, it will be instructive to yet again derive the effective potential from another point of view. We will follow \cite{ArkaniHamed:2007gg,Heidenreich:2015nta} and think of the potential as the Casimir energy density of the higher-dimensional theory integrated over the internal space,
		\begin{equation}
			V_\lab{eff}(\theta) = R\, \rho(R, \theta)\,.
		\end{equation}
		This energy density can be extracted from the stress-energy tensor's vacuum expectation value,
		\begin{equation}
			\langle T_{\mu \nu} \rangle = - \rho(R) \,\eta_{\mu \nu} - R \,\frac{\partial \rho(R)}{\partial R} \delta_{\mu}^y \delta_{\nu}^y\,. \label{eq:casimirDef}
		\end{equation}
		which, since this theory is free, can be related to the two-point function $G(x - x') = \langle \Phi(x) \bar{\Phi}(x')\rangle$ on $\mathbb{R}^{1,d} \times \lab{S}^1$,
		\begin{equation}
			\langle T_{\mu \nu} \rangle = \lim_{x' \to x} \left[\left(\partial_\mu \partial'_\nu + \partial_\nu \partial'_\mu\right) - g_{\mu \nu} \left(\partial^\rho \partial'_\rho + m^2 \right)\right] G(x - x')\,. \label{eq:setVev}
		\end{equation}
		In turn, this two-point function  can be constructed by imaging the two-point function\footnote{Since we will only evaluate this two-point function at spatial separations, we will not need its form inside the light-cone.}  in higher-dimensional flat space $\mathbb{R}^{1, d+1}$,
		\begin{equation}
			G_{\infty}(x - x') = \int\!\!\frac{\ud^{d+2} k}{(2 \pi)^{d+2}} \frac{e^{i k(x - x')}}{k^2 + m^2 - i \epsilon} = \frac{m^d}{(2 \pi)^{d/2+1}} \frac{K_{d/2}(m |x|)}{(m |x|)^{d/2}}
		\end{equation}
		around the compact space with an appropriate phase,
		\begin{equation}
			G( x- x') = \sum_{n \in \mathbb{Z}^*} e^{i n \theta} G_{\infty}(x - x' + n R \hat{y})\,. \label{eq:twoPoint}
		\end{equation}
		We exclude the $n = 0$ contribution since it will only renormalize the cosmological constant and will not affect the potential's $\theta$-dependence. In terms of the non-compact two-point function, the effective potential is
		\begin{equation}
			V_\lab{eff}(\theta) = 4 R \sum_{n \in \mathbb{Z}^*} e^{i n \theta} \left.\frac{\partial G_{\infty}(y_n^2)}{\partial y_n^2}\right|_{\mathrlap{y_n = n R \hat{y}}}
		\end{equation}
		which matches the previous result (\ref{eq:effPotExtra}).

		This form affords us a different perspective on the effective potential's discontinuous behavior. We can rewrite the two-point function (\ref{eq:twoPoint}) as
		\begin{align}
			G(x- x', y-y') = \frac{1}{R}\sum_{\ell \in \mathbb{Z}} \int\!\frac{\ud^{d+1} k}{(2 \pi)^{d+1}} \frac{e^{i k (x - x') - i m_\textsc{kk}(\theta + 2 \pi \ell)(y - y')}}{k^2 + m^2 + m_{\textsc{kk}}^2(\theta + 2 \pi \ell)^2 - i \epsilon}\,.
		\end{align} 
		via Poisson resummation.
		This expression contains the $n = 0$ piece excluded in (\ref{eq:twoPoint}), which we may remove by either explicitly subtracting or, as we will do now, by taking derivatives with respect to $\theta$. Using (\ref{eq:casimirDef}) and (\ref{eq:setVev}), we have 
			\begin{equation}
				V_\lab{eff}(\theta) \propto  \lim_{\substack{x \to x'\\ y \to y'}}\, \sum_{\ell \in \mathbb{Z}} \int\!\frac{\ud^{d+1} k}{(2 \pi)^{d+1}} \frac{k^2 \,e^{i k (x - x') - i m_\textsc{kk}(\theta + 2 \pi \ell)(y - y')}}{k^2 + m^2 + m_{\textsc{kk}}^2(\theta + 2 \pi \ell)^2 - i \epsilon}\,.
		\end{equation}
		Clearly, as the higher-dimensional mass and theta angle vanish, $m \to 0$ and $\theta \to 0$, the $\ell = 0$'s momentum space integral can become IR divergent. Indeed, 
		\begin{equation}
			\frac{1}{(2 \ell)!} \left.\frac{\ud^{2 \ell} V_\lab{eff}(\theta)}{\ud \theta^{2 \ell}}\right|_{\theta = 0} \!\propto \lim_{x \to x'} \int\!\!\frac{\ud^{d+1} k}{(2 \pi)^{d+1}} \frac{k^{2} e^{i k (x - x')}}{(k^2 + m^2)^{\ell+1}} \,.
		\end{equation}
		As $m \to 0$, we see that the integrand behaves as $k^{d - 2 \ell}$ as $k \to 0$, and so we expect trouble from this contribution when the order of the derivative exceeds $d$. This is exactly the situation we found in (\ref{eq:lowmassLimit}).

	\subsection{Other Examples} \label{sec:other}

		Our general discussion suggests that whenever an instanton expansion ``breaks down,'' we should look for a light state to blame. Perhaps a more precise way to state this is that the expansion~(\ref{eq:fourierEffPot}) cannot remain well-controlled if a state becomes almost degenerate with the vacuum. The number of these light states can be inferred from the asymptotic behavior of the expansion. We have illustrated this in two calculable examples, and it will be useful to provide evidence for this picture in others. One of the original examples of this behavior is in the Ising model \cite{Fonseca:2001dc,Langer:1967ax,Lowe:1980ez,Andreev:1964sot}, where the free energy develops a singularity close to the phase transition due to the formation of large ``droplets.'' In this section, we will discuss how this general picture applies to four-dimensional gauge theories and supersymmetric string compactifications.

		\subsubsection*{Yang-Mills, Large $N$, and Dashen's Phenomenon} \label{sec:yangMills}

			It will be useful to make more direct contact with four-dimensional gauge theories like (\ref{eq:4dGauge}) and better explain, at least qualitatively, how these theories are related to the simplified model presented in~\S\ref{sec:toy}. We want to closely mimic the discussion presented there, so we will first review the presentation of Yang-Mills theory in the Schr\"{o}dinger formalism, as this is very useful in understanding the vacuum structure of the theory. We will follow \cite{Jackiw:1977hn,Jackiw:1983nv,Jackiw:1988sf,Jackiw:2005wp}. 

			We will denote the gauge field and its field strength as $A_\mu = A_\mu^a T_a$ and $F_{\mu \nu}^a = \partial_\mu A_\nu^a - \partial_\nu A_\mu^a + f_{abc} A_\mu^b A_\nu^c$, respectively, where $T_a$ are the generators of the gauge group and $[T_a, T_b] = i f_{abc} T_c\,$. It is convenient to work in temporal gauge $A_0^a(\mb{x}) = 0$, so that the state of the system is described by the wavefunctional $\Psi[A_i^a(\mb{x}), \,\ldots\,]$, where $A_i^a(\mb{x})$ are the spatial components of the gauge field and the $\dots$ represents the wavefunctional's dependence on the other fields in the theory, like charged matter. These states evolve according to the Hamiltonian
			\begin{equation}
				\mathcal{H}_\lab{4d} = \int\!\ud^3 x\left[\frac{g^2}{2} \left(\frac{1}{i}\frac{\delta}{\delta \!A_a^i (\mb{x})} - \frac{\theta}{8 \pi^2} B_i^a\right)^2 + \frac{1}{2g^2} B_i^a (\mb{x}) B_i^a(\mb{x}) + \cdots\right], \label{eq:4dHam}
			\end{equation}
			where the $\dots$ again represents the charged matter contribution and we have introduced the chromo-magnetic field $B_i^a = -\frac{1}{2} \epsilon^{ijk} F^a_{jk}\,$. 

			We may still perform time-independent gauge transformations $A_i \to A'_i = g^{-1} A_i g + g^{-1} \partial_i g\,$ without leaving temporal gauge, so we must demand that  physical wavefunctions are invariant under such transformations,
			\begin{equation}
				\Psi[A'_i(\mb{x}) , \,\ldots\,] = \Psi[A_i(\mb{x}), \,\ldots\,]\,. \label{eq:gaugeInvariance}
			\end{equation}
			For any gauge group $\mathcal{G}$ with non-trivial homotopy $\pi_3(\mathcal{G}) = \mathbb{Z}$, like $\mathcal{G} = \lab{SU}(N)$, these gauge transformations fall into different topologically distinct classes characterized by the integer winding number
			\begin{equation}
				n = \frac{1}{24 \pi^2} \int\!\ud^3 x\, \epsilon^{ijk} \lab{tr}\left[ (g^{-1} \partial_i g)(g^{-1} \partial_j g) (g^{-1} \partial_k g)\right].
			\end{equation}
			Other groups may have a more complicated classification, but we will not consider them here.
			We can ensure that the wavefunctional is invariant under topologically trivial gauge transformations, i.e. those that are built up by iterating an infinitesimal gauge transformations, by imposing the Gauss' law constraint
			\begin{equation}
				\mathcal{D}^{ac}_i \left(\frac{1}{i} \frac{\delta \Psi}{\delta A_c^i(\mb{x})}\right) = 0
			\end{equation}
			where $\mathcal{D}^{ac}_i = \delta^{ac} \partial_i - i f^{abc} A^b_i(\mb{x})$ is the gauge-covariant derivative.

			The Hamiltonians (\ref{eq:4dHam}) is very similar to our toy model's (\ref{eq:toyHam}), at least superficially. The analog of the periodic boundary conditions (\ref{eq:toyShifts}) is the condition (\ref{eq:gaugeInvariance}) restricted to gauge transformations with unit winding. Indeed, we can remove the Hamiltonian's dependence on the theta angle by performing a transformation analogous to (\ref{eq:1dbc}), 
			\begin{equation}
				\tilde{\Psi}[A^a_i(\mb{x}), \,\ldots\,] = e^{-i \theta W[A]} \Psi[A^a_i(\mb{x}), \,\ldots\,]
			\end{equation}
			where
			\begin{equation}
				W[A_i^a] = -\frac{1}{4 \pi^2} \int\!\ud^3 x\, \epsilon^{ijk} \,\lab{tr}\!\left[\frac{1}{2} \partial_i A_j A_k + \frac{1}{3} A_i A_j A_k\right]\,.
			\end{equation}
			This term obeys
			\begin{equation}
				\frac{\delta W[A_j^b]}{\delta A^i_a(\mb{x})}  = -\frac{1}{8 \pi^2} B_i^a(\mb{x})\,,
			\end{equation}
			so after this ``gauge transformation'' the Hamiltonian simplifies to
			\begin{equation}
				\mathcal{H} = \int\!\ud^3 x\left[-\frac{g^2}{2} \frac{\delta^2}{\delta A_a^i (\mb{x})\, \delta A_a^i(\mb{x})} + \frac{1}{2 g^2} B_{i}^a (\mb{x}) B_{i}^a(\mb{x}) + \dots\right] \,, 
			\end{equation}
			analogous to (\ref{eq:crystalHam}).
			This is another way introducing of the $\theta$ parameter---it appears as a parameter in the boundary conditions we impose on wavefunctionals of the theory. Under a gauge transformation with winding number $n$, the wavefunctional transforms as
			\begin{equation}
				\tilde{\Psi}[A'^a_\mu(\mb{x}),\,\ldots\,]  = e^{-i n \theta} \tilde{\Psi}[A_\mu,\, \ldots\,]\,.
			\end{equation}
			Clearly, there is a direction in field space that corresponds to these large, topologically non-trivial gauge transformations and this direction is a circle.

			We can think of integrating out all modes besides this topologically non-trivial one to find an effective Hamiltonian in terms of the Chern-Simons number, see \cite{vanBaal:1992xj, vandenHeuvel:1994ah, Tye:2015tva,Tye:2016pxi,Bachas:2016ffl}, yielding something similar to the toy model (\ref{eq:toyHam}). This is equivalent to deriving an effective Hamiltonian for the degree of freedom that traces out a non-contractible loop in field space, though this is not an unambiguous concept \cite{Manton:1983nd,Klinkhamer:1984di,Manton:2004tk,Moore:1998swa}. The height of the potential $V(A)$ is related to the energy of the sphaleron solution, the field configuration that is the ``mountain pass'' between two vacua \cite{Manton:2004tk}, defined precisely as the minimal maximum energy along all non-contractible loops in field space. Theories with a well-defined sphaleron solution, like the electroweak sector of the Standard Model \cite{Manton:1983nd,Klinkhamer:1984di}, will have dynamics similar to the toy model with non-vanishing potential and will thus have a well-controlled instanton expansion.\footnote{The electroweak theta angle is not observable in the Standard Model \cite{Anselm:1993uj,Shifman:2017lkj}, since it can be rotated away via a $\lab{U}(1)_{B - L}$ transformation. Physical dependence can be reintroduced by explicitly adding $B+L$ violating operators~\cite{Perez:2014fja}. This is similar to how the QCD theta angle can be rotated away by a chiral transformation in the presence of massless quarks, but becomes physical away from this limit. }

			It is more difficult to derive an effective Hamiltonian for theories whose sphaleron solutions are ruled out by scaling arguments, like pure Yang-Mills or QCD. Such classically scale-invariant theories allow configurations of arbitrarily large size with arbitrarily low energy, and so the potential barrier between two topologically distinct vacua is arbitrarily small~\cite{Rubakov:2002ctg}. The instanton approximation in these theories similarly fails due arbitrarily large semi-classical configurations, and we might expect that such infrared issues signal that a gap (nearly) closes for some value of $\theta$. While we could regulate the theory by placing it on a sphere~\cite{Bachas:2016ffl}, it will be helpful to consider the dynamics of QCD via the chiral effective Lagrangian, which will allow us to identify which states are becoming light as we vary $\theta$. 

\begin{figure}
					\centering
					\includegraphics[scale=1]{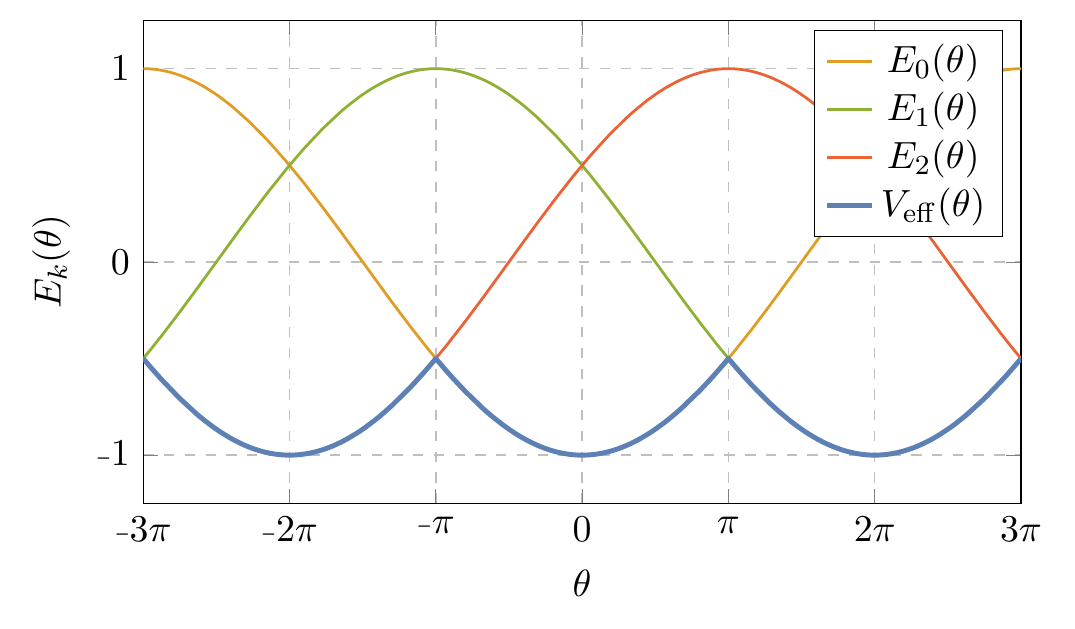}
					\caption{The energies $E_k(\theta)$ in units of $3 m \Sigma$. As $\theta \to \theta + 2\pi$, the vacua interchange among one another so that the minimum energy or effective potential is $2\pi$-periodic. \label{fig:qcdEnergies}}
			\end{figure}

			Following~\cite{Witten:1980sp,DiVecchia:1980yfw,Smilga:1998dh}, let us consider a confining gauge theory with $N$ colors and 3 fundamental quarks $\psi^f$ of equal mass $m$.\footnote{We will use $f, g= 1, 2, 3$ to denote flavor indices. See also, e.g.~\cite{Leutwyler:1992yt,Creutz:1995wf,Halperin:1998gx,Fugleberg:1998kk,Tytgat:1999yx,Vonk:2019kwv}.} This theory has a $\lab{U}(3)_\lab{L} \times \lab{U}(3)_\lab{R}$ flavor symmetry group that, at low energies, is broken to the diagonal $\lab{U}(1)_\lab{V} \times \lab{SU}(3)_\lab{V}$ by the formation of a quark condensate $2 \langle \psi_{\textsc{l}}^f \bar{\psi}_{\textsc{r}}^g \rangle = \Sigma \, \delta^{f g}$ and the chiral anomaly. The theory is then described by the sigma model
			\begin{equation}
				\mathcal{L} = -\frac{f_\pi^2}{4} \lab{tr}\big[\partial_\mu U \partial^\mu U^\dagger\big] + \Sigma \re \lab{tr} \big[\mathcal{M} \,e^{i \theta/3} U^\dagger\big] - \frac{1}{2} f_\pi^2 m_{\eta'}^2 \left(-i \log \det U\right)^2
			\end{equation}
			Here, $U$ is an element of $U(3)_\lab{V}$, $f_\pi$ is the pion decay constant, $\mathcal{M} = m \mathds{1}$ is the quark mass matrix,  and $m_{\eta'} \propto N^{-1/2}$ is the mass generated for the phase of $U$ (the ``$\eta'$ meson'') by the chiral anomaly, and we have used this anomaly to shift all $\theta$-dependence into the quark mass term. In the limit $f_\pi^2 m_{\eta'}^2 \gg m \Sigma$ so that the anomaly-induced mass term fixes $\det U = 1$, the potential energy becomes
			\begin{equation}
				-\Sigma \re \lab{tr}\big[\mathcal{M} e^{i \theta/3} U^\dagger\big] = -m \Sigma\Big[\cos\!\big(\alpha - \tfrac{\theta}{3}\big) + \cos\!\big(\beta - \tfrac{\theta}{3}\big) + \cos\!\big(\alpha + \beta + \tfrac{\theta}{3}\big)\Big],
			\end{equation}
			where we have parameterized the group element as $U = \lab{diag} \left(e^{i \alpha}, e^{i \beta}, e^{\sminus i(\alpha+\beta)}\right)$. This potential has four unique stationary points corresponding to different vacuum expectation values of $U  \propto \langle \psi^f_{\textsc{l}} \psi^g_{\textsc{r}}\rangle$: $U_0 = \mathds{1}$, $U_1 = e^{2 \pi i /3} \mathds{1}$, $U_2 = e^{\sminus 2 \pi i /3}\mathds{1}$, and ${U_{3} = \minus e^{\sminus {2 i\theta}/{3}} \, \lab{diag}\!\left(1, 1,e^{2i\theta}\right)}$, with energy densities
			\begin{equation}
				E_k(\theta) = -3 m \Sigma \,\cos\!\left(\frac{\theta - 2 \pi k}{3}\right)\,,\mathrlap{\qquad k =0, 1, 2\,\,,} \label{eq:chiralEffEnergies}
			\end{equation}
			and $E_3(\theta) = m \Sigma \cos \theta$. Since $E_3(\theta)$ is never the lowest energy, we will ignore it and the corresponding expectation value $U_3$. We plot those remaining in Figure~\ref{fig:qcdEnergies}.

			As we take $\theta \to \theta + 2\pi$, these vacua interchange among each other, $U_k \to U_{(k+1)\!\! \mod 3}$ so that the effective potential $V_\lab{eff}(\theta) = \min_k E_k$ is a periodic function of $\theta$, with Fourier coefficients
			\begin{equation}
				v_\ell = \frac{1}{2\pi} \int_{-\pi}^{\pi}\!\ud \theta \, V_\lab{eff}(\theta) e^{\sminus i \ell \theta} = \frac{3 \times 6 m \Sigma }{2 \pi \csc\big(\tfrac{\pi}{3}\big)} \frac{(-1)^\ell}{3^2 \ell^2 -1} \,.
			\end{equation}
			This situation is clearly analogous to the toy model with a $\mathbb{Z}_3$-symmetric potential discussed in \S\ref{sec:toyFrac}. As is clear from Figure~\ref{fig:qcdEnergies}, the energy gap between the states $U_0$ and $U_1$ closes at $\theta = \pi$, which in turn forces the effective potential's topological expansion to converge algebraically. This first-order phase transition spontaneously breaks CP and is known as Dashen's phenomenon~\cite{Dashen:1970et}. Its connection to the zeros of the QCD partition function has been noted before~\cite{Jackson:2001qc,Akemann:2001ir,Aguado:2002xf}.\footnote{Such techniques are more often used to understand the Roberge-Weiss \cite{Roberge:1986mm} phase transition in QCD at finite temperature and chemical potential~\cite{Barbour:1991vs,Nakamura:2013ska,Nagata:2014fra,Wakayama:2018wkc}.}

			Similar behavior appears in the $N \to \infty$ limit, where the quark mass term dominates over the anomaly. In this case, there are $\mathcal{O}(N)$ vacua labeled by the expectation value of $\gamma = -i \log \det U$ with
			\begin{equation}
				\gamma_k = 3 \pi k -  \frac{3 (-1)^k f_\pi^2 m_{\eta'}^2}{m \Sigma} \left(\theta + 3 \pi k\right) + \mathcal{O}\left(1/N^2\right),\mathrlap{\qquad k \in \mathbb{Z}\,.}
			\end{equation}
			Large $N$ pure Yang-Mills also displays this behavior. In fact, one can explicitly show~\cite{Unsal:2012zj} that, like the model of \S\ref{sec:toyFrac}, fractional instantons dominate the partition function by considering the theory compactified on $\mathbb{R}^3 \times \lab{S}^1$ with a deformation that prevents center symmetry breaking~\cite{Unsal:2008ch}. This deformed theory matches pure Yang-Mills in $\mathbb{R}^4$ up to $\mathcal{O}(1/N^2)$ corrections. That the ground state energy is a multi-branched function of $\theta$,
			\begin{equation}
				V_\lab{eff}(\theta) = \min_{k} E(\theta + 2 \pi k)\,, \label{eq:largeN}
			\end{equation}
			is a generic expectation from large $N$ counting~\cite{Witten:1980sp,Azcoiti:2003ai,Lucini:2012gg,Aitken:2018mbb}, and can be argued for by establishing the presence of a mixed 't Hooft anomaly \cite{Gaiotto:2017yup}
			 between the gauge theory's center symmetry and time reversal.

			Finally, this behavior may also be seen holographically~\cite{Witten:1998uka} by considering Type IIA string theory on $\mathbb{R}^4 \times \lab{S}^1 \times \mathbb{R}^5$ with a stack of $N$ parallel wrapped fourbranes oriented along the $\mathbb{R}^4 \times \lab{S}^1$. The theta angle arises from the integrating the Ramond-Ramond one-form along the compact cycle $\theta = \int_{\lab{S}^1} C_1$. The branes generate a background that is topologically $\mathbb{R}^4 \times \lab{D} \times \lab{S}^4$, where $\lab{D}$ is the two-dimensional disc, and we may place a two-form flux $F_2 = \ud C_1$ along this disc so that $\int_\lab{D} F_2 = \theta +2 \pi k$ with $k \in \mathbb{Z}$. It is this flux that distinguishes the different branches, similar to how the quark bilinear expectation value labeled the different branches in the chiral effective theory~(\ref{eq:chiralEffEnergies}). The standard Maxwellian kinetic energy for $F_2$ is then responsible for the ground state energy of the form
			\begin{equation}
				V_\lab{eff}(\theta) = \min_k \Lambda^4 \left(\theta + 2 \pi k\right)^2 + \mathcal{O}\big(1/N\big)\,,
			\end{equation} 
			where $\Lambda^4$ is independent of $N$ and matches the expectation~(\ref{eq:largeN}).

		\subsubsection*{Supersymmetric Compactifications}

			Finally, let us close this section by discussing some aspects of $\mathcal{N} = 1$ supersymmetric string compactifications of Type IIB string theory on Calabi-Yau orientifolds. We will follow the notation of~\cite{Baumann:2014nda,Grimm:2004uq}. For more information on this vast subject, see e.g.~\cite{Baumann:2014nda,Grimm:2004uq,Grana:2005jc,Blumenhagen:2006ci,Becker:2007zj,Denef:2008wq,Ibanez:2012zz}.

			At low energies, compactifications of Type IIB string theory along a three-dimensional Calabi-Yau orientifold can be described via four-dimensional $\mathcal{N} = 1$ supergravity theories. In particular, these theories contain a set of complex scalar fields $\phi^i$, called the K\"{a}hler coordinates on moduli space, which are described by the Lagrangian\footnote{For clarity, we have ignored the interactions between $\phi^i$ and the supergravity's gauge fields. In general, the indices $i, j$ range over the $h_+^{1,1} + h_-^{1,1} + h_-^{1,2} + 1$ moduli, where the $h^{a,b}_{\pm}$ are the even (odd) Hodge numbers of the Calabi-Yau threefold. However, in what follows we will only focus on the moduli involving the $C_4$ axion, so that~$i, j = 1, \dots, h_+^{1,1}$.} 
			\begin{equation}
				\mathcal{L} = -K_{i \bar{\jmath}}\, \partial^\mu \phi^i \es \partial_\mu \bar{\phi}^{\es \bar{\jmath}} - V_{\lab{F}}(\phi^i, \bar{\phi}^{\es \bar{\imath}})\,.
			\end{equation}
			This theory is described entirely by the holomorphic superpotential $W(\phi^i)$ and a real K\"{a}hler potential $K(\phi^i, \bar{\phi}^{\bar{\imath}})$ with the K\"{a}hler metric $K_{i \bar{\jmath}} = \partial_i \partial_{\bar{\jmath}} K$. The F-term potential 
			\begin{equation}
				V_\lab{F}(\phi^i, \bar{\phi}^{\bar{\imath}}) = e^{K} \left[K^{i \bar{\jmath}} D_i W \, \overline{D_j W} - 3 |W|^2\right]
			\end{equation}
			is a real function of the K\"{a}hler moduli, where $D_i W \equiv \partial_i W + (\partial_i K) W$, $K^{i \bar{\jmath}}$ is the inverse K\"{a}hler metric, and we work in units of the four-dimensional Planck mass, $M_\lab{pl} = 1$.

			In such string compactifications, the natural analogs of the theta angle descend from integrating a Ramond-Ramond $p$-form along a non-trivial $p$-cycle. In this section, we will focus primarily on the $C_4$ axions,
			\begin{equation}
				\vartheta_i = \frac{1}{\ell_s^4} \int_{\Sigma_4^i} C_4,
			\end{equation}
			where $\ell_s = 2 \pi \sqrt{\alpha'}$ is the string length, $i = 1, \dots, h_{+}^{1,1}$, and $\Sigma_4^i$ are a basis of the Calabi-Yau's orientifold-even four-cycles. This normalization is such that the theory is invariant under the discrete shifts $\vartheta_i \to \vartheta_i + 1$, and in order to make contact with the notation of the rest of the paper, we will define $\theta_i = 2 \pi \vartheta_i$. In the absence of vevs for the axions descending from $B_2$ and $C_2$, these $\vartheta_i$ pair with the four-cycle volumes $\tau_i$ into the complexified K\"{a}hler coordinates
			\begin{equation}
				T_i = \tau_i + i \vartheta_i\,.
			\end{equation}
			The potential is then a \emph{real} function of the $T_i$ with $W = W(T_i)$ and ${K = K(T_i, \overline{T}_{\bar{\imath}})}$. Let us focus on the case with a single K\"{a}hler modulus $T = \tau + i \vartheta$ ($\theta = 2 \pi \vartheta$), since we will not have anything to say about the interplay between multiple.
			
			Throughout this work, we have studied the partition function and corresponding effective potential as functions of the complexified theta angle.  A natural question to ask is how this complexification relates to the one imposed by supersymmetry. That is, if we shift the axion by some purely imaginary part $\theta + 2\pi i \Delta t$ (and thus move away from the origin in the $\zeta$-plane) then this seems equivalent to considering the theory at shifted four-cycle volumes $\tau - \Delta t$. If, for example, BPS Euclidean D3-branes generate the potential for $\theta$, their contributions to the superpotential come in powers of $e^{-2 \pi k \tau}$, with $k$ a positive integer. As long as this suppression persists as $k \to \infty$, we would infer that the annulus of analyticity for $\zeta = e^{i \theta}$ has outer radius $R_\lab{out} = e^{2 \pi \tau}$. Approaching $|\zeta| = R_\lab{out}$ then simply corresponds to considering the theory with vanishing~$\tau$, and so it is not shocking that light objects would appear.

			However, there is a flaw in this argument. The complexification we perform in studying the partition function's analytic structure is \emph{not} the same as the supersymmetric one, since the F-term potential is a real function of both $T$ and $\overline{T}$. Since $\overline{T} = \tau - i \vartheta$ (and not $\tau - i \bar{\vartheta}$), any shift $\vartheta \to \vartheta + i \Delta t$ is equivalent considering the anti-holomorphic components of the theory at $\tau \to \tau + \Delta t$. That is, the holomorphic and anit-holomorphic parts of the potential transform differently under such a shift, which obscures the role the volume $\tau$ play in its analytic structure. 

			Fortunately, things simplify at a supersymmetric minimum, $D_i W = 0$, where the potential becomes
			\begin{equation}
				V_\lab{F}(T, \overline{T}) = -3 \,e^{K(T, \overline{T})} \,W(T) \overline{W}(\overline{T})\,.
			\end{equation}
			This has the right structure for the above argument to work. That is, the anti-holomorphic factor $\overline{W}(\overline{T})$ does not affect the analytic structure (the location of zeros and singularities) of the holomorphic factor $W(T)$, and vice versa. If we encounter a singularity in $W(T)$ by shifting $\theta \to \theta + 2 \pi i \Delta t$, we will similarly encounter a singularity in the F-term potential. The K\"{a}hler potential could potentially remove such singularities, as it also depends on $\theta = -\pi i(T - \overline{T})$. However, at least at leading order in volumes (i.e. in the no-scale limit), it only depends on $\tau = (T+ \overline{T})/2$ and so it does not affect the F-term potential's analytic structure. So, at least for supersymmetric compactifications in the no-scale limit, this relationship between light states and the convergence of the instanton expansion is a very natural one.

\section{Quantum Gravitational Implications} \label{sec:impl}

	One of our original motivations was to better understand the physical content of the axionic Weak Gravity Conjecture. Ideally, the Weak Gravity Conjectures explain how a quantum gravitational theory reacts when we try to restore a global symmetry. However, a conjecture phrased in terms of instanton actions necessarily fails in this regard---the object the conjecture supposedly constrains becomes ill-defined exactly in the limit of interest. Instead, we have argued that the breakdown of the instanton expansion is merely a symptom of states becoming light and that it much more efficient to specify the axionic effective potential in terms of the properties of these states, rather than summing billions of Fourier harmonics.

	Can we use this perspective to reformulate the axionic Weak Gravity Conjecture,
	\begin{equation}
		S_1 \lesssim M_\lab{pl}/{f}\,, \label{eq:awgcP}
	\end{equation}
	in a way that eschews instanton actions altogether? As we discussed in Section~\ref{sec:ptFC}, the potential's Fourier coefficients exponentially decay at a rate set by the location of the singularity closest to the physical domain, denoted by $\zeta_*$. We could then define a notion of an ``asymptotic single instanton action,'' $\mathcal{S} = \log |\zeta_*|$.  This singularity is, in turn, determined by where the smallest energy gap vanishes, $\Delta E(\zeta_*) = 0$. If we assume that $\zeta_*$ is close to the unit circle, then we can solve for it in terms of information about the energy gap at physical values of $\theta$.

	For instance, if we assume that the energy gap is smallest at $\theta_*$ (so that $\zeta_* \propto e^{i \theta_*}$), then\footnote{Here, we have assumed that the eigenvalues repel simply like in (\ref{eq:energyGapToy}). If the gap behaves like $\Delta E(\zeta) \propto (\lambda^{n} + \log^n(-\zeta))^{1/n}$, we must replace $\Delta E''(\theta_*) \equiv \partial_\theta^2 \Delta E(\theta_*)$ with $\partial_\theta^n \Delta E(\theta_*)/n!$. If the gap does not have a branch point, the argument of the exponential must be multiplied by $\sqrt{2}$.}
	\begin{equation}
		|\zeta_*| \sim \exp\left(\sqrt{\frac{\Delta E(\theta_*)}{\Delta E''(\theta_*)}}\right)\,, \mathrlap{\quad\qquad |\zeta_*| \to 1\,.}
	\end{equation}
	If we interpret the aWGC (\ref{eq:awgcP}) as a statement only about the asymptotic exponential suppression of the Fourier coefficients, implicit in equations like (\ref{eq:phiEffPot}), then we could restate the conjecture as instead constraining the combination\footnote{This bound is somewhat similar to the magnetic version of the aWGC discussed in \cite{Hebecker:2017uix,Hebecker:2017wsu,Dolan:2017vmn,Reece:2018zvv}. This conjecture requires that there are strings, coupled to the axion's two-form dual, whose tensions obey~$T \lesssim f M_\lab{pl}$. In this case, there are light states that appear in the limit opposite to the one considered here, $f/M_\lab{pl} \to 0$. }
	\begin{equation}
		\sqrt{\frac{\Delta E(\theta_*)}{\Delta E''(\theta_*)}} \lesssim \frac{M_\lab{pl}}{f}\,, \label{eq:awgc2}
	\end{equation}
	in the limit $\zeta_*$ approaches the unit circle, and thus as $f/M_\lab{pl} \to \infty$.
	Unlike (\ref{eq:awgcP}), the left-hand side of this bound is clearly well-defined even in the limit it vanishes.

	We can thus unify the discussion of \cite{Banks:2003sx}, which argued that attempting to realize super-Planckian axion decay constants in string theory always forces either a state to become light or the instanton expansion to break down. We see now that these are not two separate failure modes. This also suggests another route to proving the aWGC: rather than enumerating instanton contributions, we could establish the existence of some states that quantum gravity requires become light somewhere along the axion's field space.

	We should note that neither the original formulation (\ref{eq:awgcP}) nor the reformulation (\ref{eq:awgc2}) are constraining enough to rule out theories that admit controlled super-Planckian field displacements. In the limit that $f \gg M_\lab{pl}$ and the field space becomes apparently vastly super-Planckian, the naive expectation was that the higher instanton corrections come in and ``cut down'' the distance over which the potential is smooth. However, from our general arguments, both (\ref{eq:awgcP}) and (\ref{eq:awgc2}) simply encode that there is some state (or many) coupled to the axion which becomes gapless \emph{somewhere} along the axion's field space in the limit $f/M_\lab{pl} \to \infty$. These conditions do not provide any physical information beyond this. This is effectively the content of the ``coherent instanton sum'' loophole that has been pointed out before, see e.g.~\cite{Heidenreich:2019bjd}. Where and how this gap closes would be encoded in the asymptotic behavior of the instanton phases and their subleading decay rate, respectively.

	Put another way, arguing that these unsuppressed higher instanton corrections prevent large, super-Planckian displacements by cutting down the field space assumes much more information than is contained in both (\ref{eq:awgcP}) and (\ref{eq:awgc2}). There are many seemingly smooth potentials with non-exponentially suppressed harmonics, as illustrated in Figure~\ref{fig:extra}. 

	These conjectures alone thus do not rule out the possibility of realizing large field axion inflation. If we condition on preventing such super-Planckian displacements, it is clear that the aWGC should be strengthened and it would be interesting to understand what a stronger form should look like. For instance, one may derive nontrivial constraints on axion inflation \cite{Heidenreich:2019bjd} by assuming that the instanton phases are randomly distributed. This seems to be equivalent\footnote{It is not clear whether our conclusions, which focus entirely on theories with a single axion, apply unchanged to theories with multiple. It would be interesting to understand this in future work.} to assuming that light states appear everywhere along the axion's field space as $f/M_\lab{pl} \to \infty$. Whether or not such a situation actually arises in quantum gravity remains to be seen but, even if it does, one can show \cite{Heidenreich:2019bjd} that such an assumption is not strong enough to rule out super-Planckian displacements entirely.

		\subsection{Slightly-more-natural Inflation}

			One of the best-studied inflationary potentials is that of natural inflation \cite{Freese:1990rb,Adams:1992bn},
			\begin{equation}
				V_{\textsc{ni}}(\varphi) = \Lambda^4 \left(1 - \cos \varphi/f\right)\,, \label{eq:naturalInflation}
			\end{equation}
			which can be motivated as the generic instanton-induced effective potential for an axion $\varphi$ in the limit of large instanton actions. It supports $\sim 60$ $e$-folds of large-field inflation when the axion decay constant is super-Planckian, $f \gtrsim 4 M_\lab{pl}$. We argued in \S\ref{sec:yangTopological} that potentials of the form
			\begin{equation}
				V_\alpha (\varphi) = \Lambda^4\, \re \left[\Li_{\alpha+1}\!\big(\minus\es e^{-t_* + i \varphi/f}\big)  -\Li_{\alpha+1}\!\big(\minus\es e^{-t_*}\big)\right] \label{eq:potABit}
			\end{equation}
			naturally arise in the limit that the instanton expansion becomes poorly controlled, where we have added a constant such that $V_\alpha(0) = 0$. When the singularities that define the effective potential are far from the physical domain, the potential is well-described by a single instanton action and sinusoids are the appropriate basis functions. In the opposite limit, $t_* \ll 1$, the most relevant information is the behavior of the imaginary energy gap near the physical domain (\ref{eq:epsExp}) and polylogarithms are the more appropriate basis functions (\ref{eq:polyLogExp}). Since the aWGC makes potentials like (\ref{eq:potABit}) seem to be slightly more natural than (\ref{eq:naturalInflation}) when the decay constant is super-Planckian, is natural to wonder what inflationary predictions they make.

			\begin{figure}
				\centering
				\includegraphics{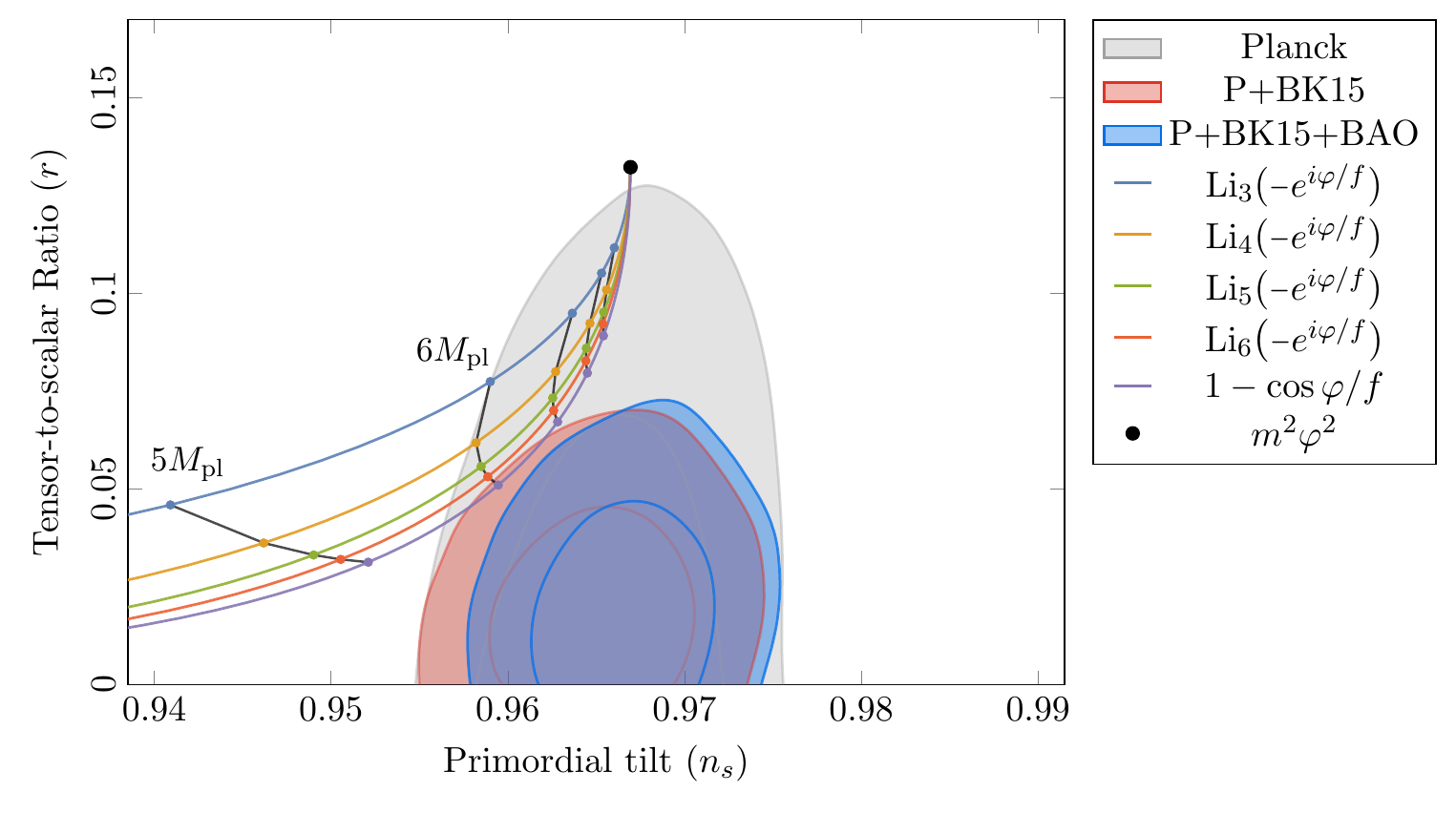}
				\caption{Predicted values of the spectral index $n_s$ and the tensor-to-scalar ratio $r$ for the potential (\ref{eq:potABit}) with $t_* = 0$ and various values of $\alpha$, natural inflation, and $m^2 \varphi^2$ at a pivot scale $N_* = 60$, compared with the \emph{Planck} 2018 $1\sigma$ and $2 \sigma$ exclusion limits. The predictions change as we vary the decay constant $f$ and we mark each curve at $f/M_\lab{pl} = 5, 6, \dots, 9$, connecting each set with a black line. As $f \to \infty$, all predictions approach that of $m^2 \varphi^2$ inflation. Similarly, the predictions from (\ref{eq:potABit}) replicate those of natural inflation when $\alpha \gg 1$. \label{fig:planckComparison}}
			\end{figure}
		
			We can study (\ref{eq:potABit})'s inflationary phenomenology in the slow-roll approximation. The slow-roll parameters $\epsilon$ and $\eta$ are defined by
			\begin{equation}
				\epsilon \equiv \frac{M_\lab{pl}^2}{2} \left(\frac{V'}{V}\right)^2 \quad \text{and} \quad \eta \equiv M_\lab{pl}^2 \es \frac{V''}{V}\,.
			\end{equation}
			We take inflation to end when $\epsilon = 1$. The number of $e$-folds of inflation between horizon crossing (at $\varphi = \varphi_*$) and the end of inflation (at $\varphi = \varphi_\lab{end})$ is given by
			\begin{equation}
				N_* \equiv \int_{\varphi_\lab{end}}^{\varphi_*} \!\frac{\ud \varphi}{M_\lab{pl}} \frac{1}{\sqrt{2 \epsilon}}\,,
			\end{equation}
			and we will assume that $N_* = 60$. We can then compute the spectral index and tensor-to-scalar ratio using
			\begin{equation}
				n_s = 1 + 2 \eta_* - 6 \epsilon_* \quad\text{and}\quad r = 16 \epsilon_*\,,
			\end{equation}
			where the $*$ indicates that we are evaluating at horizon crossing.

			In Figure~\ref{fig:planckComparison}, we compare these inflationary predictions to those of natural and $m^2 \varphi^2$ inflation, as well as the experimental constraints provided by the \emph{Planck} collaboration~\cite{Akrami:2018odb}.  Since increasing $t_*$ smoothly interpolates between (\ref{eq:potABit}) and (\ref{eq:naturalInflation}), we restrict to $t_* = 0$ and show the predictions for several different values of $\alpha$. Each model traces out a curve in the ($n_s$, $r$)-plane as we vary $f$, and all (including natural inflation) approach the predictions of $m^2 \varphi^2$ as $f/M_\lab{pl} \to \infty$. Furthermore, since (\ref{eq:potABit}) reduces to a (periodic) quadratic potential (\ref{eq:periodicQuadratic}) when $\alpha =1$ as $t_* \to 0$, its predictions are identical to that of $m^2 \varphi^2$ as long there is enough room for 60 $e$-folds of inflation to occur,~$f > 11 \sqrt{2} M_\lab{pl}$. 

			Like natural inflation, these ``slightly-more-natural'' models are in serious tension with experimental constraints---they are, after all, large-field models. However, it is interesting that making natural inflation more consistent with the aWGC by ``minimally'' introducing higher harmonics in the way suggested by the general considerations of \S\ref{sec:yangTopological} does not affect the model's ability to realize inflation, but instead makes the primordial gravitational wave signal stronger.

\section{Conclusions} \label{sec:conclusions}
	
	The main goal of this paper was to understand what goes wrong when we ``lose control of the instanton expansion.'' We have argued that the theory breaks down in the classic sense of states becoming light. Such light states induce discontinuities in the effective potential, and these discontinuities force the Fourier expansion to slowly converge. This is a simple representation of the fact that the Fourier basis is ill-suited to describe such potentials, and we used the Yang-Lee theory of phase transitions to identify an appropriate basis.

	This is not to say that the effective theory itself immediately breaks down whenever the instanton expansion does. How quickly the expansion converges depends on the behavior of the smallest energy gap in the theory \emph{everywhere} along the axion's field space, and not on the specific value of the axion. The effective theory can remain well-controlled as long as we stick far enough away from where the gap closes. We illustrated these facts in two complementary examples and described qualitatively how this picture works in some more realistic models that are not so amenable to calculation. 

	We were primarily motivated by our attempts to understand the physical content and proper formulation of the axionic Weak Gravity Conjecture, which tries to forbid the appearance of a continuous shift symmetry in the $f/M_\lab{pl} \to \infty$ limit by placing an upper bound on instanton~actions. Unfortunately, such bounds cannot have teeth---instanton actions provide little-to-no physical information about the theory when they become small, as they do in this interesting limit. We provided a well-defined reformulation (\ref{eq:awgc2}) of these conjectures in terms of the theory's smallest axion-dependent energy gap (and its derivatives). This unifies the aWGC with the others by phrasing it as quantum gravitational constraint forced on the spectrum of the theory. However, such a constraint does not prevent controlled super-Planckian displacements and must be supplemented with additional information to do so.

	In the future, it would be interesting to apply this line of thinking to theories with multiple axions, along the lines of \cite{Bestgen:1968ptz,Grossmann:1969apt}. This is analogous to understanding the energy band structure of higher-dimensional insulators and whether or not they permit exponentially-localized Wannier functions \cite{Pimentel:2019otp, Marzari:2012}. This is a relatively simple question in one dimension \cite{Kohn:1964tis,He:2001edp} that depends on the topology of the band structure in higher dimensions \cite{Brouder:2007elw}. Furthermore, it would be interesting to understand what types of phase transitions can occur along the axion's field space in consistent quantum gravitational theories. This ties into the question of whether the aWGC can be strengthened enough to forbid controlled super-Planckian displacements, and is at the heart of the swampland program. We hope to return to some of these questions in the future.

\subsection*{Acknowledgments}

	It is a pleasure to thank Tarek Anous, Daniel Baumann, Horng Sheng Chia, Alex Cole, Mehmet Demirtas, Markus Dierigl, Naomi Gendler, Jim Halverson, Ben Heidenreich, Austin Joyce, Manki Kim, Cody Long, Greg Mathys, Liam McAllister, Miguel Montero, Jakob Moritz, Guilherme Pimentel,  Matt Reece, Antonio Rotundo, Pablo Soler, Irene Valenzuela, and Thomas van Riet for many helpful conversations. We are particularly indebted to Tarek Anous, Matt Reece, and Markus Dierigl for comments on a draft. This work is supported by NASA grant 80NSSC20K0506.

	\newpage

	\appendix

	\section{The Effective Potential for General Densities} \label{app:genDen}
		
		In the main text, we found that the effective potential could be expressed as
		\begin{equation}
			V_\lab{eff}(\theta) = -\frac{1}{2 \pi} \int_{0}^{\infty}\!\ud t \, \frac{\epsilon(t)}{1 - \zeta_* e^{t - i \theta}}
		\end{equation}
		in coordinates where the imaginary energy gap vanishes at the origin, $\epsilon(0) = 0$. We may expand $\epsilon(t)$ near the origin,
		\begin{equation}
			\epsilon(t) \sim 2 \pi \sum_{\alpha} \frac{\epsilon_a t^{\alpha}}{\Gamma(\alpha+1)}\,, \mathrlap{\qquad t \to 0\,,} \label{eq:energyAsymptotics}
		\end{equation} 
		where the exponents $\alpha > 0$ are not necessarily integers, and find an expansion of the effective potential in terms of polylogarithms,
		\begin{equation}
			V_\lab{eff}(\theta) = \sum_{\alpha} \epsilon_\alpha \Li_{\alpha+1}(e^{i \theta}/\zeta_*)\,.
		\end{equation} 
		However, it may be that this expansion is not necessarily well-controlled. For instance, in \S\ref{sec:toy} we considered the density
		\begin{equation}
			\epsilon(t) = 2 \pi g^2 \sqrt{t(t + 2 t_*)}\,.
		\end{equation}
		An expansion like (\ref{eq:energyAsymptotics}) will come in powers of $t/t_*$ and necessarily break down as $t_* \to 0$. This just reflects the fact that the asymptotic behavior as $t \to 0$ changes when $t_* = 0$, compared to $t_* \neq 0$, and we expect that we can still find a useful a perturbative expansion using more sophisticated techniques. We will analyze such densities in this appendix.

		Let us consider the more general density
		\begin{equation}
			\epsilon(t) = 2 \pi \epsilon_0\, t^\alpha (t + \gamma\, t_*)^\beta
		\end{equation}
		where $\alpha, \beta, \gamma$ are arbitrary positive constants. We are interested in the behavior of
		\begin{equation}
			V_\lab{eff}(t_*, \theta) = -\epsilon_0\, t_*^{\alpha + \beta + 1} \int_{0}^{\infty}\!\ud x \, \frac{x^\alpha (x+\gamma)^\beta}{1 - \zeta_* e^{-i \theta} e^{t_* x}} \label{eq:effPotApp}
		\end{equation}
		in the $t_* \ll 1$ limit, i.e. as the density of zeros impinges upon the physical domain. Let us define
		\begin{equation}
			f(x) \equiv x^\alpha (x + \gamma)^\beta \qquad \text{and} \qquad h(x) \equiv \big[1 - \zeta_* e^{-i \theta} e^{x}\big]^{\sminus 1}\,,
		\end{equation}
		so that the effective potential takes the form of a Mellin convolution,
		\begin{equation}
			V_\lab{eff}(\theta) = - \epsilon_0\,  t_*^{\alpha + \beta+1} \int_{0}^{\infty}\!\ud x\, f(x) h(t_* x)\,.
		\end{equation}
		It will thus be extremely convenient \cite{Bleistein:1986aei,Wong:2001aai} to rewrite this as a inverse Mellin transform.

		The Mellin transform $\mathcal{M}h(z)$ of $h(x)$ is defined as
		\begin{equation}
			\mathcal{M}h(z) \equiv \int_{0}^{\infty} \!\ud x\, x^{z-1} h(x)  = -\Gamma(z) \Li_z (e^{i \theta}/\zeta_*)\,, \mathrlap{\qquad \re z > 0\,.} \label{eq:melH}
		\end{equation}
		In order to define the Mellin transform of $f(x)$, we must split it into two functions with $f_1(x) = f(x) \theta(1-x)$ and $f_2(x) = f(x) \theta(x-1)$, where $\theta(x)$ is the Heaviside step function. Their Mellin transforms can be compactly expressed in terms of the incomplete beta function 
		\begin{align}
			\mathcal{M} f_1(1-z) &= (-1)^{z-\alpha-1} \gamma^{1+\alpha+\beta - z} B_{\sminus 1/\gamma}(1+\alpha -z, 1+\beta) \nonumber \\
			\mathcal{M} f_2(1-z) &= (-1)^{1+\alpha+\beta-z} B_{\sminus \gamma}(z - \alpha -\beta - 1, 1+\beta) \label{eq:fs}
		\end{align}
		with domains of analyticity $\re z < 1 + \alpha$ and $\re z > 1 + \alpha + \beta $, respectively. It will be convenient to rewrite the second transform as
		\begin{align}
			\mathcal{M}f_2(1-z) &=  \frac{\Gamma(1 + \beta) \Gamma(z -\alpha - \beta -1)}{(-\gamma)^{z - \alpha -\beta-1} \Gamma(z - \alpha)} - \frac{(1 + \gamma)^{1 + \beta}}{1 + \beta}\,  {}_2 F_1(1, z- \alpha; \beta + 2; 1+ \gamma)  
		\end{align}
		since the hypergeometric contributions are usefully regular for all finite $z$. 

		\begin{figure}
			\centering
			\includegraphics[scale=1.05]{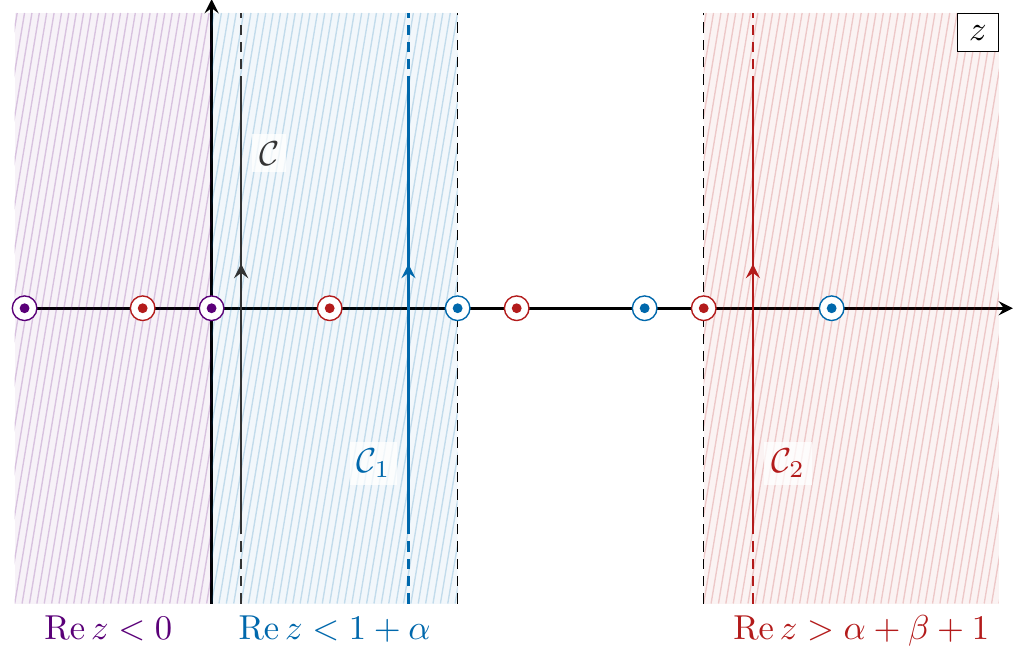}
			\caption{Regions of analyticity of the Mellin transforms (\ref{eq:melH}) and (\ref{eq:fs}) and contours of integration for the inverse transform (\ref{eq:mellin}). As we drag $\mathcal{C}_1$ and $\mathcal{C}_2$ into $\mathcal{C}$, we only pick up contributions from the poles of $\mathcal{M}f_2(1-z)$. \label{fig:mellin}} 
		\end{figure}

		The effective potential (\ref{eq:effPotApp}) may then be written as a sum over the two contributions
		\begin{align}
			V_\lab{eff}(t_*, \theta) = - \sum_{i = 1}^{2} \,\frac{\epsilon_0}{2 \pi i} \!\int_{\mathcal{C}_i} \!\ud z\,\, t_*^{-z+\alpha+\beta+1} \, \mathcal{M}f_i(1-z) \,\mathcal{M}h(z) \label{eq:mellin}
		\end{align}
		where $\mathcal{C}_1$ is a contour that runs parallel to the imaginary axis with real part $0 < c_1 < 1+ \alpha$ and $\mathcal{C}_2$ is the same but with real part $c_2 > \alpha + \beta+1$, illustrated in Figure~\ref{fig:mellin}. This representation is useful because we can derive an approximation to this integral for $t_* \ll 1$ by dragging the contour of each integral to the left, picking up the contribution from each pole we encounter. These contributions are suppressed by higher powers of $t_*$ the further leftward the pole is, and so we can systematically improve our approximation by including the contributions from 

		We now specialize to the case when $k = \alpha + \beta + 1$ is an integer. We can drag both contours to the left and combine them into a single contour $\mathcal{C}$, pictured in Figure~\ref{fig:mellin}. We  only encounter the poles of $\mathcal{M} f_2(1-z)$, allowing us to rewrite the effective potential as
		\begin{align}
			&V_\lab{eff}(t_*, \theta) = \sum_{n = 0}^{k-1} \, \res_{z = k-n} \left[\epsilon_0 (-\gamma t_*)^{k-z} \frac{\Gamma(1+ \beta)\Gamma(z-k)\Gamma(z)}{\Gamma(z - \alpha)}\Li_z(e^{i\theta}/\zeta_*)\right]  + \cdots \nonumber \\
			&=  \sum_{n = 0}^{k-1} \frac{\epsilon_0 (\gamma t_*)^{n}}{n!} \frac{\Gamma(1+\beta)\Gamma(k-n)}{\Gamma(k-n-\alpha)} \Li_{k-n}(e^{i \theta}/\zeta_*) -\frac{\epsilon_0}{2 \pi i} \int_{\mathcal{C}} \!\ud z\, t_*^{k-z} \mathcal{M}f(z) \mathcal{M}h(z) \label{eq:effPotMellinIntermediate}
		\end{align}
		where the sum of the Mellin transforms simplifies dramatically,
		\begin{align}
			\mathcal{M}f(z) &=  \mathcal{M} f_1(z) + \mathcal{M}f_2(z) = \frac{\gamma^{k - z} \Gamma(1+\alpha - z) \Gamma(z - k)}{\Gamma(\minus \beta)}\,.  \label{eq:totMellin}
		\end{align}
		Since $k$ is an integer, the poles of $\mathcal{M} f(z)$ and $\mathcal{M} h(z)$ overlap for $\re z \leq 0$ and so we find drag the contour further and further to the left, yielding the expansion
		\begin{align}
			&V_\lab{eff}(t_*, \theta) \sim \sum_{n = 0}^{k-1} \frac{\epsilon_0 (\gamma t_*)^{n}}{n!} \frac{\Gamma(1+\beta)\Gamma(k-n)}{\Gamma(k-n-\alpha)} \Li_{k-n}(e^{i \theta}/\zeta_*) \label{eq:fullPotExp}  \\
			&- \frac{\epsilon_0 (-1)^k}{\Gamma(-\beta)}\sum_{n = 0}^{\infty} \frac{(\gamma t_*)^{k+n} \Gamma(n+ \alpha + 1)}{n!\,(n+k)!} \left(\left(a_n + \log(\gamma t_*)\right)\Li_{\sminus n}(e^{i \theta}/\zeta_*) - \Li^{\scriptscriptstyle{(1)}}_{\sminus n}(e^{i \theta}/\zeta_*)\right) \nonumber
		\end{align}
		where $a_n = \psi(1+n+\alpha) - \psi(1+k+n)-\psi(n+1)$, and we have defined
		\begin{equation}
			\Li^{\scriptscriptstyle{(1)}}_{n}(z) = \frac{\partial \Li_{n}(z)}{\partial n} = -\sum_{k=2}^{\infty} \frac{\log k \, z^k}{k^n}\,.
		\end{equation}
		Note that (\ref{eq:totMellin}) vanishes for integer $\beta$, and the expansion (\ref{eq:effPotMellinIntermediate}) truncates to a finite sum of polylogarithms. This is not so surprising, as this is exactly the case when the expansion (\ref{eq:energyAsymptotics}) truncates neatly.

		We encountered two densities in the main text. The first was that of the low-dimensional model of \S\ref{sec:toy}, where the physics was primarily dominated by eigenvalue repulsion between two nearly-degenerate energy levels,
		\begin{equation}
			\epsilon(t) = 2 \pi g^2 \sqrt{t(t+2 t_*)} \quad \text{with} \quad \zeta_* = -e^{t_*}\,,
		\end{equation}
		and $t_* \sim \lambda + \mathcal{O}(\lambda^2)$.
		Using (\ref{eq:fullPotExp}), the effective potential is approximately
		\begin{align}
			&g^{\sminus 2} V_\lab{eff}(t_*, \theta) \sim \Li_{2}({e^{i \theta}}/{\zeta_*}) +  t_* \Li_{1}(e^{i \theta}/\zeta_*) \nonumber \\
			&+ \frac{t_*^2}{4} \left(\left(1 + 2 \gamma_{\textsc{e}} + 2 \log \tfrac{t_*}{2}\right) \Li_{0}(e^{i \theta}/{\zeta_*}) - 2\Li_{0}^{\scriptscriptstyle{(1)}}(e^{i \theta}/\zeta_*)\right) + \lab{c.c.} + \mathcal{O}(t_*^3)
		\end{align}
		The second was that of the extra-dimensional model presented in \S\ref{sec:extra}, where the physics was dominated by a single $(d+1)$-dimensional mode becoming gapless,
		\begin{equation}
			\epsilon(t) =  -\frac{m_{\textsc{kk}}^{d+1}}{\Gamma\big(\tfrac{d+3}{2}\big)} \left[\frac{t(t+2 m R)}{4 \pi}\right]^{\frac{d+1}{2}} \quad \text{with}\quad \zeta_* = e^{m R}\,.
		\end{equation}
		Note that, with $d = 0$, we recover the previous density (up to a sign). Furthermore, for $d =3$, the expansion (\ref{eq:fullPotExp}) truncates and we recover
		\begin{equation}
			V_\lab{eff}(t_*, \theta) = -\frac{m_{\textsc{kk}}^4}{16\pi^2}\sum_{n = 0}^{2} \frac{(2 m R)^{n}}{n!}(3-n)(4-n) \Li_{5-n}(e^{i \theta}/\zeta_*) + \lab{c.c.}
		\end{equation}
		which matches (\ref{eq:extraNatural3d}).

	\section{Asymptotic Behavior of the Thermal Partition Function} \label{app:asymp}

		In \S\ref{sec:yangTopological}, we needed to make assumptions about the asymptotic behavior of the thermal partition function in order to fully determine the effective potential from the location of the zeros and poles. This is a reflection of the simple fact that we can multiply $\mathcal{Z}_\beta(\zeta)$ by a nowhere vanishing entire function (for instance, $\exp(g(\zeta))$ with $g(\zeta)$ entire) without affecting this analytic structure. By assuming that the function $f(\zeta)$ defined in (\ref{eq:tpfF}) had vanishing order (c.f. \S\ref{sec:yangLee}), we were able to solve (\ref{eq:poincareLelong}) explicitly in terms of the zero density, and thus relate the effective potential's analytic structure to the presence of ``light'' states in the theory. We will work to justify this assumption in this appendix.

		Given the infamous complexity of nonperturbative effects in quantum field theory, we will restrain our attention to a general class of toy models. As discussed in \S\ref{sec:yangMills}, we can imagine that these toy models provide the effective low-energy description for the Chern-Simons number, for instance see \cite{Luscher:1982ma,Luscher:1983gm,vanBaal:1992xj,vanBaal:1988qm,vandenHeuvel:1994ah,Leutwyler:1992yt}. In particular, we will study the asymptotic behavior of the thermal partition function with Hamiltonian
		\begin{equation}
			\mathcal{H} = \frac{1}{2}\, h(p_A + \theta) + \lambda V(A)\,
		\end{equation}
		in the limit $|\zeta| \to +\infty$. We demand that the theory is periodic $A \sim A + 1$, take $h(p_A + \theta)$ to be a smooth function of the momentum,  and characterize the general size of the potential with the parameter $\lambda$. In order for the Hamiltonian to be bounded from below, we will require that $h(p)$ is asymptotically convex so that, in particular, $|h(p)| \to +\infty$ as $|p| \to \pm \infty$. For example, we studied this model with $h(p) = g^2 p^2$ in the main text and we will return to this case below.
	
		Since this theory is periodic with $A \sim A+1$, we can always expand this Hamiltonian using the Fourier modes $\psi_n(A) = e^{2 \pi i n A}$ and consider it as an infinite dimensional matrix
		\begin{equation}
			\mathcal{H}_{n m} = \frac{1}{2} \, h\!\left(2 \pi n + \theta\right)^2 + \lambda V_{nm}
		\end{equation}
		with
		\begin{equation}
			V_{n m} = \int_{0}^1 \!\ud A\, V(A) \, e^{2 \pi i(n - m) A}\,.
		\end{equation}
		Without loss of generality, we will shift the Hamiltonian by an overall constant so that the diagonal of the potential vanishes, $V_{nn} = 0$.

		Since we define $\zeta = e^{i \theta}$, taking $|\zeta| \to +\infty$ is equivalent to taking $\im \theta \to -\infty$. In this limit, the kinetic term dominates the potential and we may thus treat $\lambda V_{n m}$ as a small perturbation. The energy eigenvalues then behave as
		\begin{equation}
			E_{n}(\theta) \sim \frac{1}{2} h(2 \pi n + \theta) + \mathcal{O}(\lambda^2/|\theta|)\,, \mathrlap{\qquad |\zeta| \to \infty\,.}
		\end{equation}
		Intuitively, taking this limit is analogous to considering very highly excited states which are able to easily ``ignore'' the potential $V(A)$. The thermal partition function then behaves as
		\begin{equation}
			\mathcal{Z}_\beta(\theta) \sim \sum_{n \in \mathbb{Z}} e^{-\frac{1}{2} \ess \beta \ess h(2 \pi n + \theta)}\left[1 + \mathcal{O}\left(\lambda^2/{|\theta|}\right)\right] \label{eq:asympThermalPF}
		\end{equation}
		Since we defined $\mathcal{Z}_\beta(\zeta) = f(\zeta) \bar{f}(1/\zeta)$ with $f(\zeta) = \sum_{\ell=0}^{\infty} f_\ell \zeta^n$ regular about the origin, both $\mathcal{Z}_\beta(\zeta)$ and $f(\zeta)$ must have the same asymptotic behavior as $|\zeta| \to \infty$ and so we can use (\ref{eq:asympThermalPF}) to infer the behavior of $f(\zeta)$ in this limit. In particular, the large-order behavior of the coefficients $f_\ell$ must match that of the Laurent coefficients $z_\ell$ in (\ref{eq:pfLaurent}).

		If we assume that $h(p)$ grows asymptotically like $h_0 p^\alpha$ as $p \to \infty$ with $\alpha >0$, it seems reasonable to expect that $\log \mathcal{Z}_\beta(\zeta) \propto \log^\alpha \zeta$. This grows slower than $\zeta$ for any $\alpha < \infty$ and thus the partition function should have vanishing order. However, it is unclear if the sum over $n$ ruins this expectation and we will have to work a bit to see that this does not happen. We can use Poisson resummation to express (\ref{eq:asympThermalPF}) as a Laurent series
		\begin{equation}
			\mathcal{Z}_\beta(\zeta) = \sum_{\ell \in \mathbb{Z}} z _\ell \zeta^\ell + \mathcal{O}(\lambda^2/|\theta|)
		\end{equation}
		with Laurent coefficients
		\begin{equation}
			z_\ell = \int_{\mathbb{R}}\!\ud n\, e^{-2 \pi i n \ell} e^{-\frac{1}{2} \beta h(2 \pi n)}\, \label{eq:laurentFT}
		\end{equation}
		If the partition function is entire with vanishing order, then these Laurent coefficients satisfy
		\begin{equation}
			\limsup_{\ell \to +\infty} \frac{\ell \log \ell}{\log(1/|f_\ell|)} = \limsup_{\ell \to +\infty} \frac{\ell \log \ell}{\log(1/|z_\ell|)} = 0\,,\label{eq:flOrder}
		\end{equation}
		which is equivalent to characterizing the asymptotic behavior of the Fourier transform (\ref{eq:laurentFT}).

		Let us assume that $h(p) = \sum_{k = 0}^{2m} h_k p^k$ is a polynomial\footnote{We can extend this theorem to entire functions $h(p)$ that satisfy $h'(\infty) = -h'(-\infty) = \infty$ and whose real part can be bounded from below by $\re h(p) \geq h(a \re p) - h(b \im p) - C$ for $p \in \mathbb{C}$ and positive constants $a$, $b$, and $C$. Furthermore, using the results of \cite{Knockaert:2012ubf}, we can further relax this assumption to convex functions $h(p)$ that are analytic within a strip $|\im p\,| < c$ for $c > 0$. For ease of presentation, though, we will restrict ourselves to simple even-degree polynomials in the text.} of even degree, with $h_{2 m} > 0$ and $m \in \mathbb{Z}_+$. Then, a theorem by \cite{Chung:1998fte} bounds the Fourier transform (\ref{eq:laurentFT})
		\begin{equation}
			|z_\ell| \leq C e^{-\epsilon h^*(\ell)}
		\end{equation}
		by the Legendre transform
		\begin{equation}
			h^*(\ell) = \sup_{p \in \mathbb{R}} \left( p \ess \ell - \frac{1}{2} \es \beta \es h(p)\right)\,,
		\end{equation} 
		for some positive constants $\epsilon$ and $C$. This implies that
		\begin{equation}
			\log 1/|z_\ell| \geq \epsilon' \ell^{\frac{2m}{2m -1}} (\label{eq:fourierCoeffBound})
		\end{equation}
		with $\epsilon'$ a positive constant. As long as $m < \infty$, (i.e. as long as the energy asymptotically grows slower than exponentially), we find that the order vanishes (\ref{eq:flOrder}) and the partition function is determined entirely by its zeros and singularities. That is, we do not need to specify any extra data at the boundary $|\zeta| \to \infty$ to solve the Poisson-like equation (\ref{eq:poincareLelong}), and we can safely set $g(\zeta)$ to a constant in (\ref{eq:poincareSol}).

		We can explicitly confirm this behavior in the toy model (\ref{eq:toyHam}) where $h(p) = g^2 p^2/2$ and $m =1$. We argued above that the potential can be ignored in the $|\zeta| \to \infty$ limit, and so the partition function reduces to the Jacobi theta function (\ref{eq:jacobiTriple}). This can be explicitly factorized as   
		\begin{equation}
		\mathcal{Z}_{\beta}(\zeta) =(q^2, q^2)_\infty\,  (\minus q \zeta, q^2)_\infty \, (\minus q \zeta^{\sminus 1}, q^2)_\infty\,,
		\end{equation}
		where
		\begin{equation}
			(a, q)_{n} = \prod_{\ell = 0}^{n} \left(1 - a q^\ell \right)\,
		\end{equation}
		is the $q$-Pochhammer symbol, so that we can identify $f(\zeta) = (q^2, q^2)_\infty^{1/2} (-q \zeta, q^2)_\infty$. This has the series expansion
		\begin{equation}
			f(\zeta) = (q^2, q^2)_\infty^{1/2} \sum_{\ell = 0}^{\infty} \frac{q^{\ell^2}}{(q^2, q^2)_\ell} \zeta^\ell\,.
		\end{equation}
		For large $k$, the denominator approaches a constant
		\begin{equation}
			(q^2, q^2)_\infty \sim \sqrt{\frac{2 \pi}{\log(1/q)}} \exp\left(\frac{\pi^2}{6 \log q} -\frac{1}{24} \log q\right), \mathrlap{\qquad q \to 1\,,}
		\end{equation}
		and so 
		\begin{equation}
			\log 1/|z_\ell| \sim \log 1/|f_\ell| \sim \ell^2 \log q\,, \mathrlap{\qquad \ell \to \infty} 
		\end{equation}
		as predicted by (\ref{eq:fourierCoeffBound}). Thus, as is well-known~\cite{NIST:DLMF}, this Jacobi theta function is an entire function in $\zeta$ with vanishing order.

\newpage
\phantomsection
\addcontentsline{toc}{section}{References}
\bibliographystyle{utphys}
\bibliography{axionwgc}

\end{document}